\documentclass[12pt]{article}
\usepackage{epsfig,amsfonts,amssymb}
\usepackage{graphicx}
\usepackage{epsfig}
\usepackage{hyperref}
\usepackage{cite}
\def\ZZZ{{\hbox{ Z\kern-1.6mm Z}}}
\def\RRR{{\hbox{ R\kern-2.4mm R}}}
\def\CCC{{\hbox{ C\kern-2.0mm C}}}
\def\zzz{{\hbox{z\kern-1mm z}}}

\newcommand{\nn}{\nonumber \\}

\newcommand{\vt}{\vartheta}

\newcommand{\qeq}{{\hbox{=\kern-2.3mm ? \kern.5mm }}}
\renewcommand{\qeq}{=}

\newcommand{\eps}{\epsilon}

\newcommand{\vp}{\varphi}

\newcommand{\MM}{{\cal M}}

\newcommand{\OO}{{\cal O}}

\newcommand{\wt}{\widetilde}

\newcommand{\NN}{{\cal N}}

\newcommand{\be}{\begin{equation}}
\newcommand{\ee}{\end{equation}}
\newcommand{\ben}{\begin{eqnarray}\displaystyle}
\newcommand{\een}{\end{eqnarray}}

\newcommand{\refb}[1]{(\ref{#1})}
\newcommand{\p}{\partial}
\newcommand{\sectiono}[1]{\section{#1}\setcounter{equation}{0}}

\def\one{{\hbox{ 1\kern-.8mm l}}}
\def\zero{{\hbox{ 0\kern-1.5mm 0}}}

\newcommand{\bea}[1]{\begin{eqnarray}\label{#1} }
\newcommand{\eea}{\end{eqnarray}}

\begin{document}

\baselineskip 24pt

\begin{center}
{\Large \bf  S-duality Improved Superstring Perturbation Theory}

\end{center}

\vskip .6cm
\medskip

\vspace*{4.0ex}

\baselineskip=18pt

\centerline{\large \rm Ashoke Sen}

\vspace*{4.0ex}

\centerline{\large \it Harish-Chandra Research Institute}
\centerline{\large \it  Chhatnag Road, Jhusi,
Allahabad 211019, India}

\vspace*{1.0ex}
\centerline{\small E-mail:  sen@mri.ernet.in}

\vspace*{5.0ex}

\centerline{\bf Abstract} \bigskip

Strong - weak coupling duality in string theory allows us to compute physical
quantities both at the weak coupling end and at the strong coupling end. 
Furthermore perturbative
string theory can be used to compute corrections to the leading
order formula  at both ends. We explore the possibility of 
constructing a smooth interpolating formula
that agrees with the perturbation expansion at both ends and
 leads to a fairly accurate determination of the 
quantity in consideration over the entire range of the coupling constant. 
We apply this to study the mass of the stable non-BPS state
in SO(32) heterotic / type I string theory with encouraging results. In
particular our result suggests that after taking into account
one loop corrections to the mass in the heterotic and type I string theory, the
interpolating function determines the mass within 10\% accuracy 
over the entire range of coupling
constant. 

\vfill \eject

\baselineskip=18pt

\tableofcontents

\sectiono{Introduction} \label{s1}

At present we do not have a fully non-perturbative definition of string theory
except in some special backgrounds involving AdS spaces. 
As a result we can only compute perturbative corrections
around a given background (see \cite{1209.5461} for up to date results on superstring
perturbation theory)
which is expected to break down at finite or strong 
coupling. S-duality provides a way out at strong coupling by mapping the problem
to a weak coupling problem in a dual string theory. However there are no known
techniques for systematic computation at finite coupling.

Certain supersymmetric quantities can be computed for all values of the coupling since
once they are computed at weak coupling they remain valid at all couplings. This
includes spectrum of BPS states, certain terms in the low energy effective action ({\it e.g.} the
prepotential in type II string theory compactified on a Calabi-Yau 3-fold) etc.
Recently remarkable progress has been made towards determining certain other class
of terms
in the low energy effective action whose form is not protected against quantum corrections, but
are sufficiently constrained by supersymmetry so that by knowing the perturbative answer
at various ends we can completely fix these terms
(see {\it e.g.} \cite{1004.0163} and references therein). However this still leaves
open the question of how to determine the wide class of other observables whose form
is not in any way restricted by supersymmetry. Most of the interesting observables like
the S-matrix and masses of non-BPS states fall in this category.

In this paper we explore the possibility that knowing the behaviour at strong
and weak coupling and matching the results from two ends, we may be able to
get fairly
accurate results for physical quantities even at finite coupling. The idea is as follows.
Let us denote by $F^W_m(g)$ and $F^S_n(g)$ respectively 
the contributions  to a given physical quantity up to $m$-th
order at the weak coupling end
($g\to 0$) and $n$-th order at the strong coupling end ($g\to\infty$). 
We then try to look for a smooth interpolating function $F_{m,n}(g)$
whose Taylor series expansions at the
weak and the strong coupling ends match those of the functions $F^W_m(g)$ and
$F^S_n(g)$ to appropriate order. Under favourable circumstances 
the function $F_{m,n}(g)$ may come reasonably
close to the actual function $F(g)$ for sufficiently large $m,n$.
Since the
perturbation expansion in string theory is an asymptotic expansion, we do not expect
that we can approach arbitrarily close to the exact result; but
the question is whether we can reach fairly close to the exact result following
this procedure.\footnote{In a different context but similar spirit, ref.\cite{0706.1555} attepmted
to find an approximate formula for the negative mode eigenvalue of the Schwarzschild
black hole  as a function
of dimension $D$ using the known behaviour at large and small $D-3$.}

We shall apply this procedure to study the mass of the lightest SO(32)
spinor state in SO(32) heterotic or equivalently type I string theory. Due to charge
conservation this state is guaranteed to be stable even though it breaks all
supersymmetry. In SO(32)
heterotic string theory this is a perturbative string state\cite{het,het1,het2} whereas in type I string
theory this is described by a stable non-BPS 
D0-brane\cite{9808141,9809111} (see also
\cite{9803194,9805019,9806155,9810188}). 
Thus it is meaningful to look for a function $F(g)$ that will give
the mass of this state as a function of the string coupling constant $g$.

The rest of the paper is organised as follows. In \S\ref{s2} we fix our conventions,
describe our strategy for finding the interpolating function and also explicitly
find the interpolating function at the leading order. In \S\ref{s3} we compute
first subleading correction to the mass of the stable non-BPS state in
type I string theory. In \S\ref{s4} we find the first subleading correction to the mass of
stable non-BPS state in SO(32)
heterotic string theory. In \S\ref{sinterpol} 
we find the interpolating function taking into account the subleading corrections at the two
ends and compare the result with the leading order interpolating function, as well as the 
strong and weak coupling expansions. We find close matching of all these functions
within about 10\%, indicating that already at this order the interpolating function may be
within 10\% of the exact result for all values of the coupling. In \S\ref{snew}
we discuss the results obtained using other interpolation methods and find that all such
methods give results within 10\%\ of the results of \S\ref{sinterpol}.
In appendix \ref{s5} we test the efficiency of our interpolation algorithm by applying
it on several test functions. Appendix \ref{sa} contains some technical details 
of the analysis carried out in \S\ref{s4}.

\sectiono{Conventions and Strategy}  \label{s2}

We begin by fixing the various normalization conventions we shall be using
in our analysis. We denote by $G_{H\mu\nu}$ and $G_{I\mu\nu}$ the heterotic
and type I metric, defined so that the fundamental strings in the respective string
theories have tension $1/2\pi$ in these metrics. We shall choose the dilatons $\phi_H$
and $\phi_I$ in the two theories so that the part of the action involving the metric,
dilaton and the SO(32) gauge fields take the form:
\be \label{e1het}
S_H=\int d^{10} x e^{- 2 \phi_H} \sqrt{\det G_H}\left[ {1\over 2} R_H + 2 G_H^{\mu\nu} 
\p_\mu\phi_H \p_\nu \phi_H - {1\over 16} G_H^{\mu\nu} G_H^{\rho\sigma}
Tr_V (F_{\mu\rho} F_{\nu\sigma})\right]
\ee
for the heterotic string theory and
\be \label{e1type}
S_I=\int d^{10} x \sqrt{\det G_I}\left[ e^{- 2 \phi_I}  \left\{ {1\over 2} R_I + 2 G_I^{\mu\nu} 
\p_\mu\phi_I \p_\nu \phi_I \right\} -  C  e^{-\phi_I} 
G_I^{\mu\nu} G_I^{\rho\sigma}
Tr_V (F_{\mu\rho} F_{\nu\sigma})\right]
\ee
for the dual type I string theory. Here $Tr_V$ denotes trace in the vector representation of
SO(32), 
\be \label{ecvalue}
C=2^{-13/2} \pi^{-7/2} 
\ee
and $R_H$ and $R_I$ denotes the scalar curvatures computed from the 
heterotic and type I metrics respectively.
Note that the overall normalization of the terms involving the metric and the dilaton
can be changed by shifting $\phi_H$ and $\phi_I$, but once these terms have been
fixed the normalization of the gauge field kinetic term is no longer arbitrary. For the
heterotic string theory this normalization was determined in \cite{het2} while for
type I string theory this can be found {\it e.g.} in \cite{9510169}.\footnote{In \cite{9510169} the
action was written as a 8+1 dimensional integral in the T-dual type I$'$ description.
Here we have expressed the
action as a 9+1 dimensional integral in type I description.}

We now introduce the Einstein metric $g_{\mu\nu}$ via the field redefinitions
\be \label{ehcan}
G_{H\mu\nu} = e^{\phi_H/2} g_{\mu\nu}, \quad G_{I\mu\nu} = e^{\phi_I/2} g_{\mu\nu}\, .
\ee
In terms of the metric $g_{\mu\nu}$ the action takes the form
\be \label{eican}
S_H=\int d^{10} x \sqrt{\det g}\left[ {1\over 2} R -{1\over 4} g^{\mu\nu} 
\p_\mu\phi_H \p_\nu \phi_H - {1\over 16} e^{-\phi_H/2} g^{\mu\nu} g^{\rho\sigma}
Tr_V (F_{\mu\rho} F_{\nu\sigma})\right]
\ee
and
\be \label{extracan}
S_I=\int d^{10} x \sqrt{\det g}\left[ {1\over 2} R -{1\over 4} g^{\mu\nu} 
\p_\mu\phi_I \p_\nu \phi_I -C  e^{\phi_I/2} g^{\mu\nu} g^{\rho\sigma}
Tr_V (F_{\mu\rho} F_{\nu\sigma})\right]\, .
\ee
Comparing \refb{eican} and \refb{extracan} we see that the two actions agree if we
make the identification
\be  \label{ephirel}
e^{(\phi_H+\phi_I) /2} =  (16 C)^{-1}\, .
\ee
We shall define the heterotic coupling $g_H$ and the type I coupling $g_I$ via the
relations
\be
g_H \equiv e^{\langle \phi_H\rangle}, \quad g_I = e^{\langle \phi_I\rangle}\, .
\ee
Eqs.\refb{ephirel}, \refb{ecvalue}  then give
\be \label{ehirel}
g_H g_I = 2^{-8} C^{-2} = 2^{5} \pi^{7}\, .
\ee

To test the duality between the heterotic and type I string theory, we can
compare the fundamental heterotic string tension with the type I D-string tension.
Heterotic string tension in Einstein frame is given by
\be
T_H = {1\over 2\pi} (g_H)^{1/2}\, .
\ee
On the other hand the
type I D-string tension in Einstein frame is given by\cite{9510017,9602052}
\be
T_D = {1\over 2} (2\pi)^{5/2} (g_I)^{-1/2}\, .
\ee
Note that there is an extra factor of $1/\sqrt 2$ compared to that of the tension
of the type IIB D-string to take into account the effect of orientifold projection.
Using \refb{ehirel} we see that $T_D=T_H$, in agreement with the heterotic - type
I duality\cite{9503124} that identifies the fundamental heterotic string with the type I D-string.

{}From now on all the masses will be given in the Einstein metric unless mentioned
otherwise. In the convention described above,
the mass of the lightest SO(32) spinor state in the heterotic string theory is given by
\be \label{eweak}
2\, (g_H)^{1/4} (1 + \OO(g_H^2)) \, .
\ee
On the other hand the mass of the stable non-BPS D0-brane in type I string theory,
transforming in the spinor representation of SO(32), can be computed as follows.
To leading order in $g_I$, 
the D0-brane is connected by marginal deformation to a D-string anti-D-string pair
wrapped on a compact circle of radius $1/\sqrt 2$ measured in the type I 
metric\cite{9808141}. Thus its mass, measured in the type I metric, will be $2\pi \sqrt 2$
times the tension of a D-string. Converting this to the Einstein metric the mass of the
D0-brane is given by
\be \label{estrong}
2 \sqrt 2 \, \pi \, T_D  (g_I)^{-1/4}  ( 1 + \OO(g_I)) = 2^3 \,  \pi^{7/2} \, (g_I)^{-3/4} (1 + \OO(g_I)) 
= 2^{-3/4}\, \pi^{-7/4} \, (g_H)^{3/4} (1 + \OO(g_H^{-1}))\, .
\ee
Comparing \refb{estrong} and \refb{eweak} we see that the leading order heterotic
and type I results meet at  $g_H= (2\pi)^{7/2}$. In view of this we introduce a
rescaled coupling parameter $g$ via
\be \label{edefg}
g_H = 2^{7/2} \pi^{7/2} g, \quad g_I = 2^{3/2} \pi^{7/2} g^{-1}\, ,
\ee
so that the two formul\ae\ meet at $g=1$. 
Furthermore since at the meeting point the mass is given by $2^{15/8} \pi^{7/8}$,
we defined a renormalized mass function $F(g)$ via the relation
\be \label{edefF}
M(g) = 2^{15/8} \pi^{7/8} F(g)\, .
\ee
In terms of $g$ the leading order weak and the strong
coupling formul\ae\ \refb{eweak} and \refb{estrong} 
for the renormalized mass function $F(g)$ can be expressed as
\be  \label{eleading}
F^W_0(g) = g^{1/4}, \qquad F^S_0(g) = g^{3/4}\, .
\ee
Our goal will be to explore to what extent we can determine the full function $F(g)$ by
finding an interpolating function 
that matches onto the above functions (and perturbative corrections to them)
at the two ends.

Let us denote by $F^W_m(g)$ and $F^S_n(g)$ the formul\ae\ for $F(g)$
in the
weak and strong coupling limits to $m$-th and $n$-th order in expansion in powers
of $g$ and $g^{-1}$ respectively.
In that case we shall choose our interpolating function as\footnote{Clearly many other
interpolations are possible. In particular we could use 
(fractional) power of a rational function for this purpose. In each case
we need to determine the efficiency of the interpolating algorithm by studying its
convergence properties. As we shall see, for the problem of studying the mass of
lightest SO(32) spinor states in heterotic string theory, \refb{einterpol} seems to give
reasonable results.}
\be \label{einterpol}
F_{m,n}(g) = g^{1/4} \bigg[1 + a_1 g +\cdots a_m g^{m} +
b_{n}g^{m+1} + b_{n-1} g^{m+2} + \cdots + b_1 g^{m+n} 
+  
g^{m+n+1}\bigg]^{1/\{2(m+n+1)\}}\, .
\ee
By construction this formula reduces to the correct forms given in \refb{eweak} and
\refb{estrong} in the weak and strong coupling limits. The coefficients $a_1,\cdots a_m$
are determined by demanding that the expansion of $F_{m,n}$ in powers of $g$ matches
the weak coupling perturbation expansion to $m$-th order, while the coefficients
$b_1,\cdots b_n$ are determined by demanding that the expansion of 
$F_{m,n}$ in powers of $1/g$ matches
the strong coupling perturbation expansion to $n$-th order.\footnote{Note that the
$a_k$'s and $b_k$'s which appear in the analysis of each $F_{m,n}$ are different
and have to be determined afresh every time.}
Since in the heterotic string theory the expansion is actually in powers of $g^2$, this
will imply vanishing of the $a_m$'s for odd $m$. 

Note that this procedure is not
foolproof since the term inside the square bracket in \refb{einterpol}
could become negative and hence
the right hand side of \refb{einterpol} will cease to give a real function. This will signal
 breakdown of this procedure. 
However for sufficiently smooth functions we could  expect such negative $a_k$ and/or
$b_k$ coefficients
to be small even if they are present. In this case the term inside the square bracket in
\refb{einterpol} will remain positive for all positive $g$ and the 
procedure should continue to work.
Nevertheless
this approach will clearly be insensitive to terms in $F(g)$ whose
 Taylor series expansion vanishes at both ends, {\it e.g.} $e^{-A g - B/g}$ times any
 polynomial in $g, g^{-1}$ for positive constants $A$, $B$. More generally since the
 perturbation expansion in string theory is expected to represent asymptotic series at
 both ends, we do not expect to get arbitrarily close to the exact result by going to
 arbitrarily high order. Typically for any given value of $g$ 
 the best result will be reached at some particular order in the perturbation theory.
 The hope is that this approach may take us sufficiently close to the exact formula
 over the entire range of $g$.
In appendix \ref{s5} we have tested this procedure on several test functions.

  \begin{figure}
\begin{center}
\epsfysize=5cm
\epsfbox{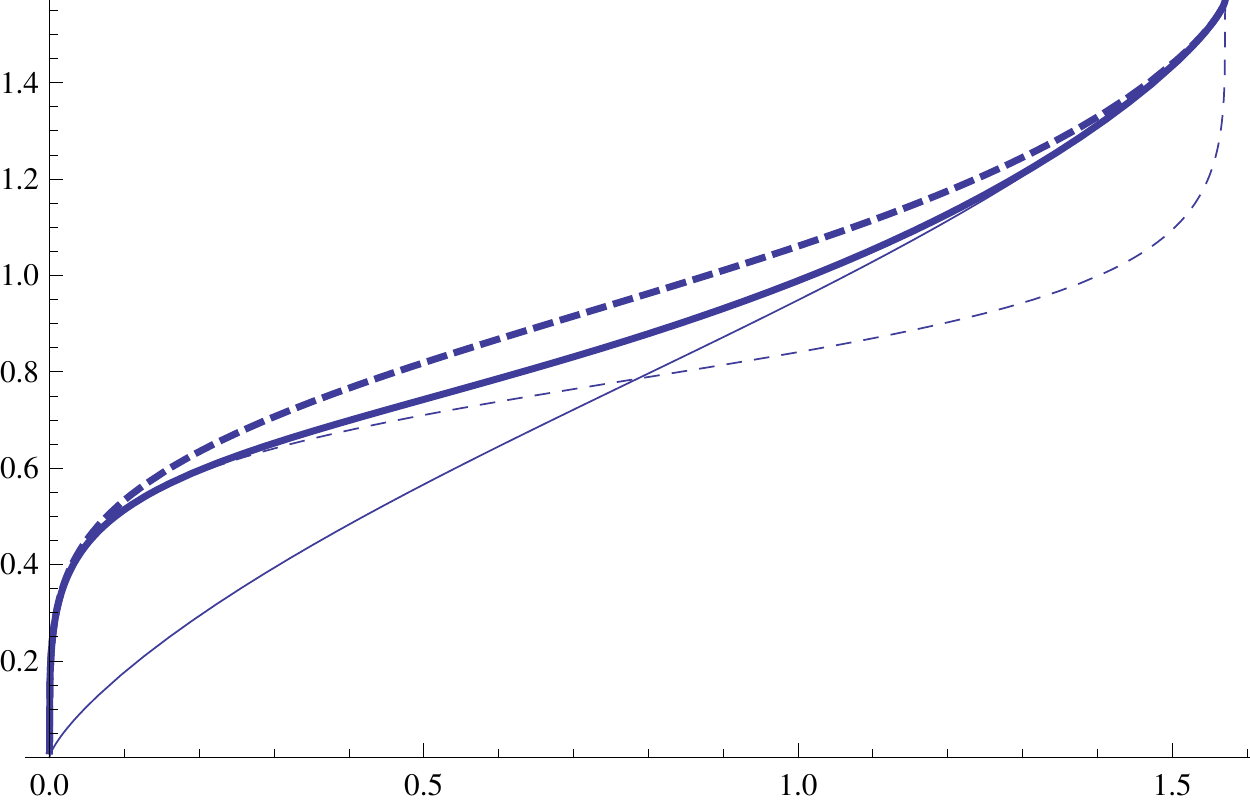}
\end{center}
\caption{Graph of $\tan^{-1}F(g)$ vs. $\tan^{-1} g$  for $F=F^S_0$ (thin solid curve),
$F^W_0$ (thin dashed curve) and the interpolating functions $F_{0,0}$ (thick dashed curve) and
$F_{1,0}$ (thick solid curve).  This should be compared with Fig.~\ref{f3} which describes
similar curves after taking into account the first subleading corrections at both ends.
\label{f1}}
\end{figure}
For example for the leading order functions \refb{eleading} we have the interpolating function
\be
F_{0,0}(g) = g^{1/4} (1 + g)^{1/2}\, .
\ee
Also knowing that the weak coupling expansion begins at order $g^2$ so that $a_1=0$, we get
\be
F_{1,0}(g) = g^{1/4} (1 + g^2)^{1/4}\, .
\ee
In Fig.~\ref{f1} we have plotted the functions $F^S_0(g)$, $F^W_0(g)=F^W_1(g)$ as well
as $F_{1,0}(g)$ and $F_{0,0}(g)$.  

\sectiono{Strong coupling expansion} \label{s3}

\renewcommand{\wt}{\tilde}

We shall now compute $F^S_1(g)$ by determining 
the first order correction $\Delta M$
to the
mass formula from the strong coupling end, \i.e.\ in type I string theory.
This is given by open string one loop correction to the energy of the non-BPS
D0-brane of type I string theory,  and can be expressed as\cite{9510017}\footnote{The
variable $s$ is related to the variable $t$ appearing in eq.(9) of \cite{9510017} as
$t=4\pi s$. There is a factor of 1/2 in our expression compared to that in \cite{9510017}
since we are considering the self energy of the D0-brane, and hence we do not get a factor
of 2 in the spectrum by exchanging the two ends of the string. The explicit form of the 
open string partition function is of course different since we have a non-BPS D0-brane in
type I string theory while \cite{9510017} was considering BPS D-p-branes in type II string theory.}
\be \label{edeltam}
-\Delta \, M
=  {1\over 2} \, g_I^{1/4} \, (8\pi^2)^{-1/2} \, \int_0^\infty \, s^{-3/2} \, ds \, 
\left[ Z_{NS;D0D0} - Z_{R;D0D0} + Z_{NS;D0D9}-Z_{R;D0D9}\right]
\ee
where $Z_{NS;D0D0}$, $Z_{R;D0D0}$, $Z_{NS;D0D9}$, $Z_{R;D0D9}$ denote 
respectively the contributions from the NS and R sector open strings with both ends
on the D0-brane and NS and R sector open strings with one end on the D0-brane
and the other end on the $D9$-brane.
Explicit calculation gives\cite{9903123,0003022,0012167}
\ben
Z_{NS;D0D0} &=& {1\over 2} {f_3(\wt q)^8\over f_1(\wt q)^8}
+ 2^{5/2} (1-i) {f_3(i\wt q)^9 f_1(i\wt q) \over f_2(i\wt q)^9 f_4(i\wt q)}
- 2^{5/2} (1+i) {f_4(i\wt q)^9 f_1(i\wt q) \over f_2(i\wt q)^9 f_3(i\wt q)}\, , \nonumber \\
Z_{R;D0D0} &=& {1\over 2} {f_2(\wt q)^8 \over f_1(\wt q)^8}\, , \nonumber \\
Z_{NS;D0D9} &=& 16\sqrt 2\, {f_1(\wt q) f_2(\wt q)^9\over f_4(\wt q)^9 f_3(\wt q)}\, , \nonumber \\
Z_{R;D0D9} &=& 16\sqrt 2\, {f_1(\wt q) f_3(\wt q)^9\over f_4(\wt q)^9 f_2(\wt q)}\, , 
\nonumber \\
\een
where
\be
\wt q \equiv e^{-\pi s}\, , 
\ee
\ben
f_1(q) &\equiv& q^{1/12} \prod_{n=1}^\infty (1-q^{2n}) = \eta(2\tau)\, ,  \quad 
q\equiv e^{2\pi i \tau}\nonumber \\
f_2(q) &\equiv& \sqrt 2\, q^{1/12} \prod_{n=1}^\infty (1+q^{2n}) =\sqrt 2\, 
 \eta(4\tau) / \eta(2\tau)\, , \nonumber \\
f_3(q) &\equiv& q^{-1/24} \prod_{n=1}^\infty (1+q^{2n-1}) =\eta(2\tau)^2/ (\eta(4\tau)\eta(\tau) )\, , 
\nonumber \\
f_4(q) &\equiv& q^{-1/24} \prod_{n=1}^\infty (1-q^{2n-1}) = \eta(\tau)/\eta(2\tau)\, .\nonumber \\
\een

Individual terms in \refb{edeltam} are both infrared (IR) and ultraviolet (UV) 
divergent; so we need
to be careful about using the right prescription for the IR and UV regularization.
Since $2\pi s$ denotes the proper length of the open string progagator flowing in the loop
in the Schwinger representation, 
the IR regularization is done by putting a uniform upper cut-off $\Lambda$ on $s$ in all
the integrals. Regulating the UV divergence is more subtle.
As we shall describe shortly, instead of putting a 
uniform lower cut-off on the integrals, the correct prescription is to use a lower 
cut-off $\eps$
on all the integrals involving $Z_{R;D0D0}$, $Z_{NS,D0D9}$, $Z_{R;D0D9}$ and
the first term in $Z_{NS;D0D0}$, but use a cut-off $\epsilon/4$ on the integrals
involving the second and third terms of $Z_{NS;D0D0}$. Thus the integral appears as
\be  \label{efirst}
-\Delta \, M
=  - \wt K_s\, (g_I)^{1/4}\, 
\ee
where
\ben \label{edefks}
 \wt K_s 
&\equiv & - {1\over 2}\, 
(8\pi^2)^{-1/2}  \, \lim_{\Lambda\to\infty} \lim_{\epsilon\to 0} \bigg[
\int_\epsilon^\Lambda \, s^{-3/2} \, ds \, \bigg\{ {1\over 2} {f_3(\wt q)^8\over f_1(\wt q)^8}
- {1\over 2} {f_2(\wt q)^8 \over f_1(\wt q)^8} \nonumber \\ &&
+ 16\sqrt 2\, {f_1(\wt q) f_2(\wt q)^9\over f_4(\wt q)^9 f_3(\wt q)}
- 16\sqrt 2\, {f_1(\wt q) f_3(\wt q)^9\over f_4(\wt q)^9 f_2(\wt q)}\bigg\}\nonumber \\
&& + \int_{\epsilon/4}^\Lambda \, s^{-3/2} \, ds \, 
\bigg\{2^{5/2} (1-i) {f_3(i\wt q)^9 f_1(i\wt q) \over f_2(i\wt q)^9 f_4(i\wt q)}
- 2^{5/2} (1+i) {f_4(i\wt q)^9 f_1(i\wt q) \over f_2(i\wt q)^9 f_3(i\wt q)}\bigg\}\bigg]
\, .
\een
Using known modular transformation laws
of the $f_i$'s we can also express \refb{edefks} in the `closed string channel':
\be \label{eclosedpre}
 \wt K_s =-  \,  \lim_{\Lambda\to\infty} \lim_{\epsilon\to 0} 
 {1\over 4\pi} (8\pi^2)^{-1/2} \bigg[
\, \int_{\pi/\Lambda}^{\pi/\eps} dt (C_{00}
+C_{09} + C_{09}^*)  +  \int_{\pi/4\Lambda}^{\pi/\eps} dt  (\MM+\MM^*)\bigg]
\, ,
\ee
where
\ben \label{eclosed}
C_{00} &=& \left({\pi\over t}\right)^{9/2} \left[ {f_3(q)^8\over f_1(q)^8} - {f_4(q)^8\over
f_1(q)^8}
\right]\, ,\nonumber \\
\MM &=& 2^{9/2} \left[ {f_3(iq)^9 f_1(iq)\over f_2(iq)^9 f_4(iq)}
- {f_4(iq)^9 f_1(iq)\over f_2(iq)^9 f_3(iq)}
\right]\, , \nonumber \\
C_{09} &=& 2^{9/2} \left[ {f_4(q)^9 f_1(q)\over f_2(q)^9 f_3(q)}
- {f_3(q)^9 f_1(q)\over f_2(q)^9 f_4(q)}
\right]\, , \nonumber \\
q &\equiv& e^{-t}\, .
\een
In going from \refb{edefks} to \refb{eclosedpre} we have made a change of variables
$t=\pi/s$ in the first integral and $t=\pi/4s$ in the second integral, explaining the limits
of integration in \refb{eclosedpre}.
Physically $C_{00}$ denotes the cylinder amplitude with both boundaries lying on the
D0-brane, given by the inner product between the boundary states of the D0-brane,
$\MM$ denotes the Mobius strip amplitude with the boundary lying on the D0-brane,
given by the inner product between the boundary states of the D0-brane and the crosscap,
and $C_{09}$ is the cylinder amplitude with one boundary on the D0-brane and the other
boundary on the D9-brane, given by the inner product between the boundary states of
the D0-brane and the D9-brane.
In \refb{eclosedpre} the parameter $t$ has the interpretation of the proper length of the closed
string propagator in the Schwinger representation, and hence the upper cut-off on
$t$ should be a uniform number for all the integrals. This is indeed the case in
\refb{eclosedpre} since all the upper cut-offs are $\pi/\eps$, but this required choosing
the lower cut-off on the $s$ integral in precisely the way we have chosen in 
\refb{edefks}.

\begin{figure}
\begin{center}
\epsfysize=5cm
\epsfbox{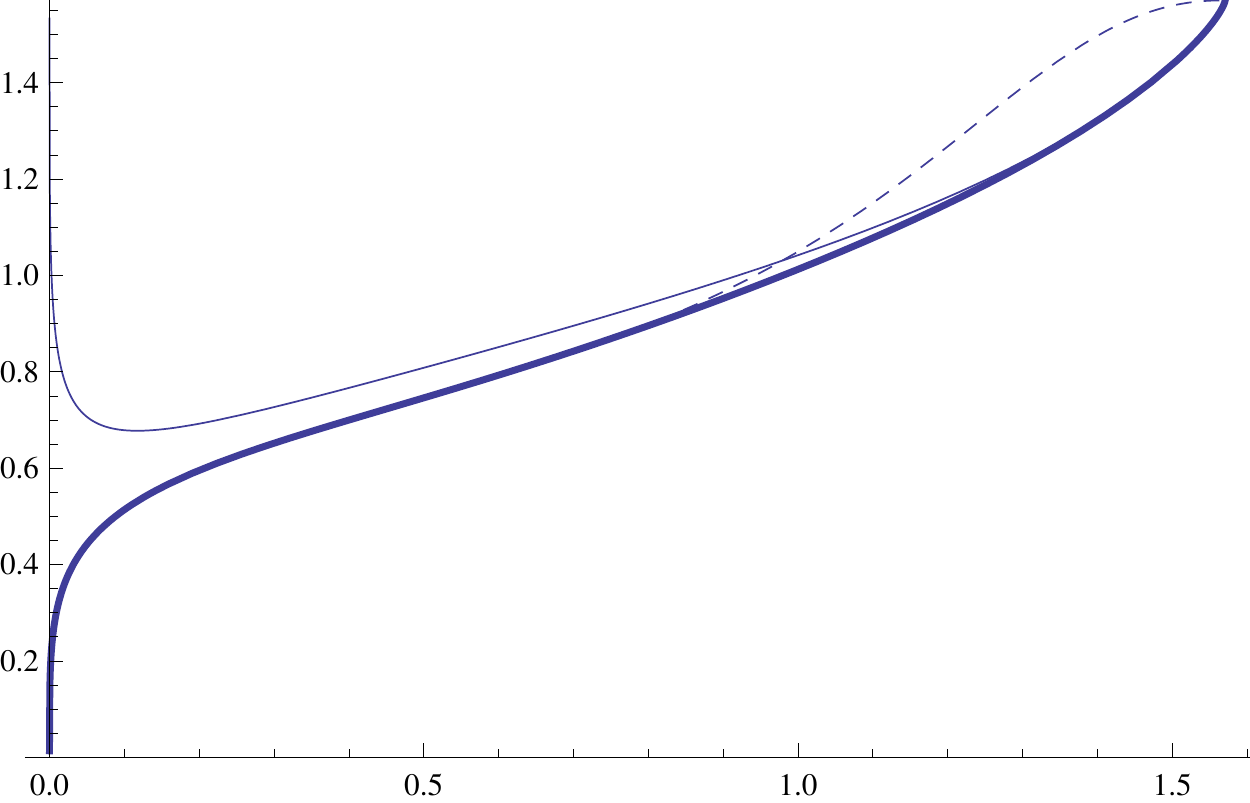}
\end{center}
\caption{Graph of $\tan^{-1}F(g)$ vs. $\tan^{-1} g$  for $F=F^S_1$ (thin solid curve),
$F=F^W_2(=F^W_3)$ (thin dashed curve) and the interpolating function
$F=F_{3,1}$ (the thick solid curve). \label{f3}}
\end{figure}

Expressing \refb{efirst} in terms of $g$ defined in \refb{edefg}, carrying out the rescaling 
given in \refb{edefF}, and combining this with the leading order strong coupling result given in 
\refb{eleading} we see that the strong coupling result for $F(g)$ corrected to first order in $1/g$
is given by
\be 
F^S_1(g) = g^{3/4}\, (1 + K_s g^{-1}), \qquad K_s \equiv 2^{-3/2}  \wt K_s\, .
\ee
Numerical evaluation
of the integrals appearing in \refb{edefks} gives
\be \label{eksresult}
K_s \simeq .351\, .
\ee
A graph of $\tan^{-1}F^S_1(g)$ vs. $\tan^{-1}g$ can be found in
Fig.~\ref{f3}, but
we shall postpone the analysis of this function till \S\ref{sinterpol} by which time
we shall also compute
the first subleading corrections from the weak coupling end.

\sectiono{Weak coupling expansion} \label{s4}

Next we shall compute the first order correction to $M$ in the weakly coupled heterotic
string theory leading to an expression for $F^W_2(g)=F^W_3(g)$. The left-handed part of the state
is created by the spin field of SO(32) acting on the vacuum. The right handed part
can be chosen in different ways. In the light cone gauge Neveu-Schwarz-Ramond (NSR)
formalism the NS sector states can be taken to be of the form\cite{het1}
\be
\psi^i_{-3/2} |0\rangle, \quad \psi^i_{-1/2}  \psi^j_{-1/2} \psi^k_{-1/2} |0\rangle, \quad
\psi^i_{-1/2} \alpha^j_{-1}|0\rangle
\ee
where $\psi^i_{-n}$ and $\alpha^i_{-n}$ are transverse fermionic and bosonic oscillators 
respectively with $1\le i\le 8$. This gives a total of 8+56+64=128 states. Together with the
128 fermionic states arising from the R-sector,
these form a single long supermultiplet transforming in the spinor
representation of the gauge group SO(32). Thus all the states suffer the same mass renormalization
and we can focus on any one of them for computing the correction to $M$.
We shall use covariant NSR formulation\cite{FMS} for our computation and choose the 
unintegrated vertex
operator in the $-1$ picture to be
\be
i\, c \, \bar c\, e^{-\varphi} \bar S_\alpha(\bar z) \psi^1(z) \psi^2(z) \psi^3(z) e^{i k_0 X^0}
\ee
where $\varphi$ is the bosonized ghost of the $\beta$-$\gamma$ system\cite{FMS}, $b$, $\bar b$,
$c$, $\bar c$ are the diffeomorphism ghosts, $\bar S_\alpha$ are the spin fields of SO(32) and
$k^0$ is the energy of the particle in the rest frame given by $M$. 
The unintegrated vertex operator in the zero picture is then given by
\ben \label{evertex}
c\bar c V_0(k,z)= i\, c \, \bar c\,  e^{i k_0 X^0} \, \bar S_\alpha(\bar z)\,  \bigg\{\psi^1(z) \psi^2(z) \p X^3(z)
+  \psi^2(z) \psi^3(z) \p X^1(z)  \nonumber \\  + \psi^3(z) \psi^1(z) \p X^2(z)
+ i k_0 \psi^1(z) \psi^2(z) \psi^3(z) \psi^0(z)\bigg\}\, , 
\een
plus terms of order $e^\vp$ whose correlation functions vanish by
$\vp$-charge conservation. For computing the torus amplitude we need to convert one of the
vertex operators to integrated vertex operator $\wt V_0$ by removing the $c\bar c$ factor. 
If $\delta M^2$ denotes the one loop
correction to $M^2$, then up to an overall multiplicative factor $\delta M^2$ is
given by 
\be \label{edeltam2}
\delta M^2 \sim (g_H)^2 \, \int {d^2\tau} \int {d^2 z} \, \langle V_0(-k, 0) \wt V_0(k, z) 
\rangle_{matter} Z_{ghost}\, ,
\ee
where $\tau$ is the modular
parameter of the torus, integrated over the fundamental domain, 
$\langle~\rangle_{matter}$ denotes  correlation function in the 
matter sector on the torus multiplied by the matter partition function and $Z_{ghost}$ denotes the
ghost partition function after removal of ghost zero modes.

To evaluate this contribution, let us first focus on the contribution from the
holomorphic fermions $\psi^\mu(z)$. The possibly non-vanishing correlation functions are of two
types: $\langle \psi^1(0) \psi^2(0) \psi^1(z) \psi^2(z)\rangle$ and its permutations
and $\langle \psi^1(0) \psi^2(0) \psi^3(0) \psi^0(0) \psi^1(z) \psi^2(z)
\psi^3(z) \psi^0(z)\rangle$. To evaluate these we first perform a double Wick
rotation on $\psi^0$ to make it $\psi^4(0)$ corresponding to some Euclidean direction
4, at the cost of picking up a factor of $i$. Next we introduce complex fermions
$\chi^k$ via $\chi^k = (\psi^k + i\, \psi^{k+4})/\sqrt 2$ so that we can express $\psi^k$ as
$(\chi^k +\bar\chi^k)/\sqrt 2$ for $1\le k \le 4$. Then we have
\ben
&& \langle \psi^1(0) \psi^2(0) \psi^1(z) \psi^2(z)\rangle 
= - \langle \psi^1(0) \psi^1(z) \psi^2(0) \psi^2(z)\rangle \nonumber \\
&& = -{1\over 4} \bigg(\langle \chi^1(0) \bar\chi^1(z) \chi^2(0) \bar\chi^2(z)\rangle
+ \langle \bar\chi^1(0) \chi^1(z) \chi^2(0) \bar\chi^2(z)\rangle \nonumber \\
&& +\langle \chi^1(0) \bar\chi^1(z) \bar\chi^2(0) \chi^2(z)\rangle
+\langle\bar\chi^1(0)  \chi^1(z) \bar\chi^2(0) \chi^2(z)\rangle\bigg)\, .
\een
Now for a single complex fermion like $\chi^1(z)$ the correlation function on the
torus is given by, up to a phase,
\be \label{ecorr}
\langle \chi^1(0) \bar\chi^1(z) \rangle = (\eta(\tau))^{-1} \, {\vt_{11}'(0)\over \vt_{11}(z)}
\vt_\nu (z), \quad  \langle \bar\chi^1(0) \chi^1(z) \rangle = (\eta(\tau))^{-1} \, 
{\vt_{11}'(0)\over \vt_{11}(z)}
\vt_\nu (-z)\, ,
\ee
where $\nu$ denotes the spin structure on the torus taking values 00, 01, 10 and 11 and 
$\vt_\nu$ are the Jacobi theta functions. After combining the contribution from all the
holomorphic fermions and the superconformal ghosts, and taking into account the
$\tau$-dependent normalization factors the result is, up to an overall constant factor,
\ben \label{eh1}
 && \langle \psi^1(0) \psi^2(0) \psi^1(z) \psi^2(z)\rangle\nonumber \\ 
&=& -{1\over 4} (\eta(\tau))^{-4} \left({\vt_{11}'(0)\over \vt_{11}(z)}\right)^2  
{1\over 2} \, \sum_\nu \delta_\nu \bigg( 
\vt_\nu(z)^2 \vt_\nu (0)^2 + 2 \vt_\nu(z)
\vt_\nu(-z) \vt_\nu (0)^2 + \vt_\nu(-z)^2 \vt_\nu (0)^2\bigg)\, , \nonumber \\ 
\een
where $\delta_\nu=1$ for $\nu=11$ and 00  and $\delta_\nu=-1$ for $\nu=10$ and $01$.
The extra factor of 1/2 in \refb{eh1} comes from the GSO projection in the right-moving
(holomorphic) sector leading to the sum over spin structures $\nu$.
Using the Riemann identity
\ben \label{eriemann}
\sum_\nu \delta_\nu \, \vt_\nu(z_1) \vt_\nu(z_2) \vt_\nu(z_3) \vt_\nu(z_4) &=&
 2 \vt_{11} ((z_1+z_2+z_3+z_4)/2) \vt_{11} ((z_1+z_2-z_3-z_4)/2) \nonumber \\ &&
\vt_{11} ((z_1-z_2-z_3+z_4)/2) \vt_{11} ((z_1-z_2+z_3-z_4)/2)\, , \nn
\een
and that $\vt_{11}(0)=0$ we see that the right hand side of \refb{eh1} vanishes.

Thus 
we are left with the two point function of the operators appearing in the last term inside
the curly bracket in
\refb{evertex}:
\be \label{emaster}
 (k_0)^2 \langle e^{-i k_0 X^0(0)} e^{i k_0 X^0(z)} S^\alpha (0) 
S_\alpha(\bar z)   \psi^1(0) \psi^2(0) \psi^3(0) \psi^4(0) \psi^1(z) \psi^2(z)
\psi^3(z) \psi^4(z)\rangle_{matter} \, Z_{ghost}
\ee
where we have taken into account an extra $-$ sign from the Wick rotation taking
$\psi^0$ to $i \psi^4$.
Now following the same method as described above, the net contribution from the ten
holomorphic fermions and the superconformal ghosts to \refb{emaster}
is given by
\ben \label{e1pre}
&& {1\over 2^4} (\eta(\tau))^{-4} \left({\vt_{11}'(0)\over \vt_{11}(z)}\right)^4  \,
{1\over 2} \, \sum_\nu \delta_\nu\bigg(
\vt_\nu(z)^4 + 4\vt_\nu (z)^3 \vt_\nu(-z) + 6 \vt_\nu(z)^2 \vt_\nu(-z)^2 \nonumber \\
&&\qquad \qquad \qquad \qquad + 
4\vt_\nu (z) \vt_\nu(-z)^3 + \vt_\nu(-z)^4\bigg)\, .
\een
We could manipulate this further using the Riemann identity, but for reasons that
will become clear in appendix \ref{sa}  we shall postpone this till the end.

The rest of the contribution can also be evaluated using standard method. Using the known
correlator between the spin fields\cite{atickas1} we get the contribution from the
32 anti-holomorphic fermions to be
\be \label{e2}
\langle \bar S^\alpha(0) \bar S_\beta(\bar z) \rangle = {1\over 2} \, \delta_{\alpha\beta} \, (\overline{\eta(\tau)})^{-16} 
\left( {\overline{\vt_{11}'(0)}\over \overline{\vt_{11}(z)}}\right)^4\sum_{\nu'}  \overline{\vt_{\nu'}
(z/2)^{16}}
\, ,
\ee
where the factor of 1/2 now arises from the GSO projection on the left-handed
(anti-holomorphic) fermions.
Finally the contribution from the ten scalars corresponding to the space-time coordinates
and the diffeomorphism ghosts together give, up to a normalization,
\be \label{e3}
(\tau_2)^{-5} {1\over |\eta(\tau)|^{16}} \exp[-{4\pi \, z_2^2 / \tau_2}] 
\left| {\vt_{11}(z)\over \vt_{11}'(0)}\right|^4\, ,
\ee
where we have used the on-shell condition $(k_0)^2 = 4$.
Substituting \refb{e1pre}, \refb{e2}, \refb{e3} into \refb{emaster}, and using the 
known relation $\vt_{11}'(0) = -2\pi \eta(\tau)^3$, 
we can now write down the general formula for $\delta M^2$:
\ben \label{egendm}
&& \delta M^2 = {1\over 64} \, \NN\, (g_H)^2 \, (k_0)^2 \, 
\int d^2 \tau \int d^2 z\, \bigg[ 
\bigg\{ (\overline{\eta(\tau))^{-4} \vt_{11}(z)^{-4}}
\, \sum_{\nu'}  \overline{\vt_{\nu'}(z/2)^{16}}\bigg\}
 \nonumber \\
&& \bigg\{ \sum_\nu \delta_\nu\bigg(
\vt_\nu(z)^4 + 4\vt_\nu (z)^3 \vt_\nu(-z) + 6 \vt_\nu(z)^2 \vt_\nu(-z)^2  + 
4\vt_\nu (z) \vt_\nu(-z)^3 + \vt_\nu(-z)^4\bigg) \nonumber \\
&&((\eta(\tau))^8 (\vt_{11}(z))^{-4}  \bigg\}  \bigg\{ (\eta(\tau))^{-14} (\overline{\eta(\tau)})^{-14}  
(\vt_{11}(z))^2 (\overline{\vt_{11}(z)})^2 \, 
\exp[-{4\pi \, z_2^2 / \tau_2}] \, (\tau_2)^{-5}\bigg\}\bigg] \nn
\een
where $\NN$ is an overall normalization constant.
In this expression the factor inside the first curly bracket 
gives the contribution from the 32 left-moving fermions,
the factor inside the second curly bracket 
gives the contribution from the 10 right-moving fermions and the
commuting superconformal ghosts, and the factor inside the third curly bracket
gives the contribution from the 
ten scalars and the diffeomorphism ghosts. The normalization factor $\NN$ 
has been computed in appendix
\ref{sa} by comparing the result with the expected result in the effective field theory.
The result is
\be \label{envalue}
\NN = {1\over 2^{10} \pi^8}\, .
\ee
Using the fact that all the $\vt_\nu$ except $\vt_{11}$ are even under $z\to -z$, and
$\vt_{11}(z)$ is odd under $z\to -z$ we can express the term in the second line of
\refb{egendm} as
\be \label{e1}
\left(16 \sum_\nu \delta_\nu
\vt_\nu(z)^4 - 16 \vt_{11}(z)^4\right)
= - 16\, \vt_{11}(z)^4\, ,
\ee
where in the last step we have again made use of the Riemann identity \refb{eriemann}.
Finally using the fact that
$k_0^2 = M^2$ to leading order, and using \refb{edefg}, we get
\be \label{ehfin}
\delta M =  M \, K_w\,  g^2\, ,
\ee
where
\be \label{ekw}
K_w =-{1\over 64 \, \pi} \, 
\int d^2 \tau \int d^2 z\, \sum_{\nu'}  \left\{\overline{\vt_{\nu'}(z/2)^{16}}\right\}
(\overline{\eta(\tau)})^{-18} (\eta(\tau))^{-6}
\left( {\vt_{11}(z)\over \overline{\vt_{11}(z)}}\right)^2
\exp[-{4\pi \, z_2^2 / \tau_2}] \, (\tau_2)^{-5}\bigg]\, .
\ee
It is easy to verify that the integrand is invariant under $z\to z+1$, $z\to z+\tau$
and $\tau\to \tau+1$ and $(z,\tau)\to (z/\tau, -1/\tau)$. The domain of integration over $z\equiv z_1+iz_2$ 
and $\tau\equiv \tau_1+i\tau_2$
is the fundamental region which can be taken to be $0\le z_1<1$, $0\le z_2< \tau_2$,
and the regin in $\tau$ plane bounded by the curves $\tau_1=\pm 1/2$ and $|\tau|=1$.
Numerical evaluation of the integral gives
\be \label{ekwvalue}
K_w \simeq .23
\, .
\ee
Using \refb{edefg}, \refb{edefF},
the weak coupling expansion of the mass formula $F(g)$ up to order $g^3$ now takes the form
\be
F^W_2(g)=F^W_3(g) = g^{1/4}\, (1 + K_w g^2)\, .
\ee

\sectiono{Analysis of the results and interpolating function} \label{sinterpol}

Let us first summarize the results of \S\ref{s3} and \S\ref{s4}. We have found that the weak
and strong coupling expansions of the mass function $F(g)$ are given by
\be 
F^W_2(g) = g^{1/4} (1 + K_w g^2 + \OO(g^4)), \quad
F^S_1(g) = g^{3/4} (1 + K_s g^{-1} + \OO(g^{-2}), \quad
K_w \simeq .23, \quad K_s \simeq .351\, .
\ee
Note that both $K_s$  and $K_w$ are smaller
than unity. This implies that the corrections from both ends remain smaller than unity at $g=1$
where the leading order weak and the strong coupling curves meet. Given this it is not 
unreasonable to expect that string perturbation theory may be able to
give a fairly good result for the function $F(g)$ over the entire range of $g$.

\begin{figure}
\begin{center}
\epsfysize=5cm
\epsfbox{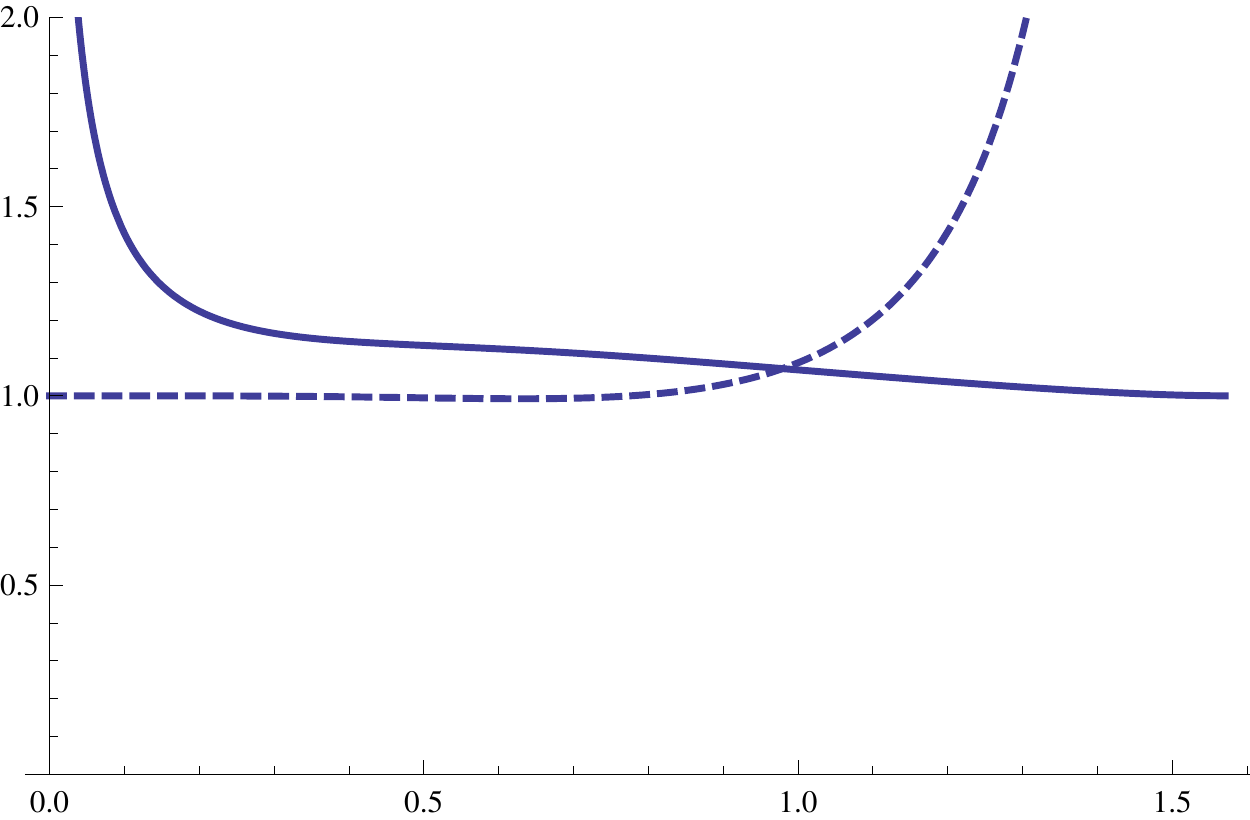}
\end{center}
\caption{Graph of $F^W_2(g)/F_{3,1}(g)$ (dashed curve) and
$F^S_1(g)/F_{3,1}(g)$ (continuous curve) vs. $\tan^{-1}g$.
\label{f4}}
\end{figure}

We shall now follow the procedure outlined in \S\ref{s2} to find the interpolating functions
whose Taylor series expansion around $g=0$ and/or $g=\infty$ agree with those of
$F^W(g)$ and/or $F^S(g)$.
We find
\be
F_{0,1}(g)= g^{1/4}\, (1 + 4\,  K_s g +g^2)^{1/4}\, ,
\ee
\be 
F_{1,1}(g) =g^{1/4}\, (1 + 6\,  K_s g^2 +g^3)^{1/6}\, ,
\ee
\be 
F_{2,0}(g) =  g^{1/4}\, (1 + 6\, K_w g^2 +g^3)^{1/6}\, ,
\ee
\be 
F_{2,1}(g) =  g^{1/4}\, (1 + 8\, K_w g^2 + 8 K_s g^3 +g^4)^{1/8}\, ,
\ee
\be 
F_{3,0}(g) =  g^{1/4}\, (1 + 8\, K_w g^2 +g^4)^{1/8}\, ,
\ee
and 
\be
F_{3,1}(g) = g^{1/4}\, \left(1 + 10\, K_w g^2 +10 \, K_s  g^4 +g^5\right)^{1/10}\, .
\ee
Fig.~\ref{f3} shows the weak and strong coupling results
$F^W_2$ and $F^S_1$ including the first subleading corrections and the interpolating function
$F_{3,1}(g)$.
Fig.\ref{f4} shows the ratios $F^W_2(g)/F_{3,1}(g)$ and
$F^S_1(g)/F_{3,1}(g)$ as a function of $\tan^{-1}g$. We see that these ratios remain close
to unity during most of the range of $g$ except at very large and small values of $g$
where $F^W_2(g)$ and $F^S_1(g)$ respectively are clearly bad approximations to the
actual function $F(g)$. More specifically, if we denote by $g_c$ the point where
$F^W_2(g)$ and $F^S_1(g)$ meet, then $F^W_2(g)$ agrees with $F_{3,1}(g)$ below $g_c$ 
within 8\%
and $F^S_1(g)$ agrees with $F_{3,1}(g)$ above $g_c$ within 8\%.

\begin{figure}
\begin{center}
\epsfysize=5cm
\epsfbox{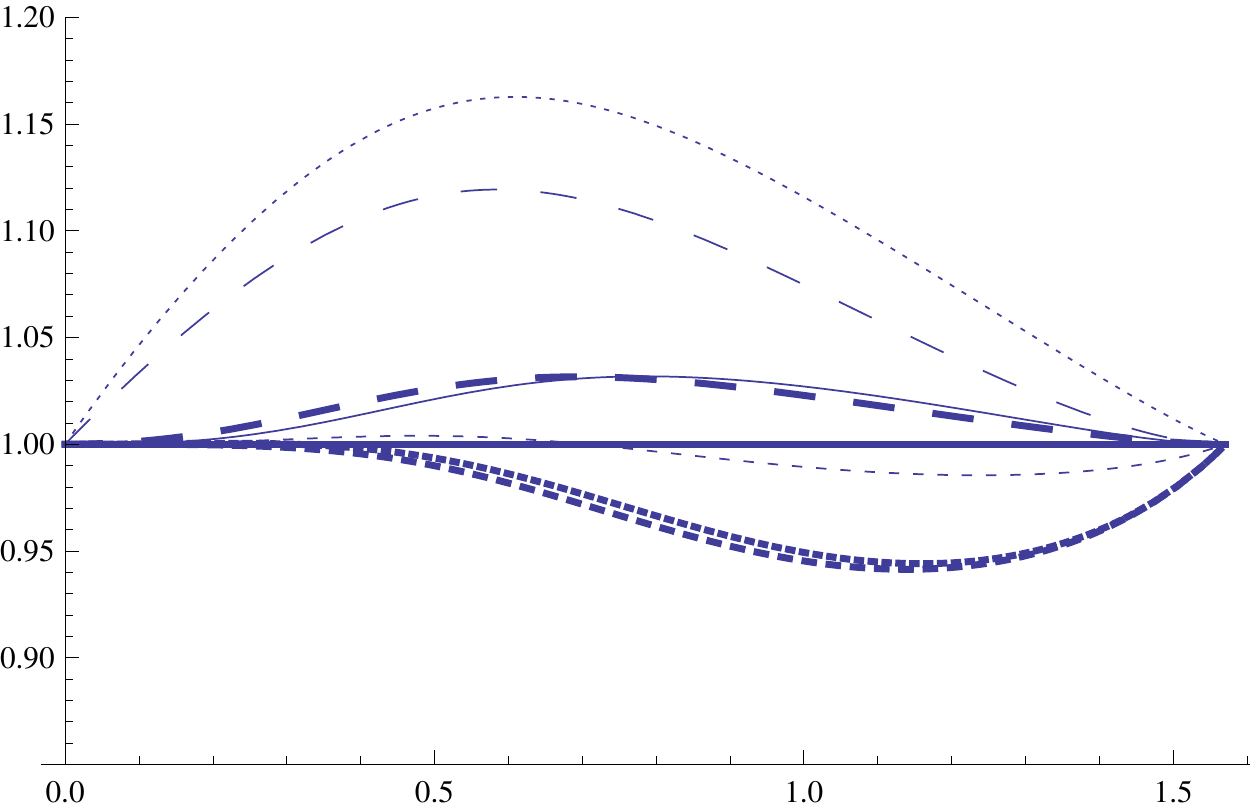}
\end{center}
\caption{Graph of $F_{m,n}(g)/F_{3,1}(g)$ vs. $\tan^{-1}g$ for various $(m,n)$. 
The labels are as follows:  thin dots for $F_{0,0}$,  thick dots for $F_{1,0}$,
small thin dashes for $F_{2,0}$, small thick dashes for $F_{3,0}$, large thin dashes for
$F_{0,1}$, large thick dashes for $F_{1,1}$, continuous thin line for $F_{2,1}$ and
continuous thick line for $F_{3,1}$.
\label{f2}}
\end{figure}

To estimate how close we are to the actual
function $F(g)$, we can analyze how the interpolating functions at different orders
differ from each other. To test this we have plotted in Fig.~\ref{f2} the ratio of
$F_{m,n}(g)/F_{3,1}(g)$  for $0\le m\le 3$, $0\le n\le 1$
as a function of
$\tan^{-1} g$. As we can see, most of 
the ratios remain within 10\% of unity throughout the whole range
of $g$, indicating that $F_{3,1}(g)$ may be within 10\% of the actual function.
In particular we see that  $F_{3,1}$ and $F_{2,1}$
lie within 5\% of each other
over the entire range of $g$.  

\begin{table}
\begin{center} \def\st{\vrule height 3ex width 0ex}
\begin{tabular}{|l|l|l|l|l|l|l|l|l|l|l|} \hline
$(m,n)$ & (0,0) & (0,1) & (1,0) & (1,1) & (2,0) & (2,1) & (3,0) & (3,1)
\st\\[1ex] \hline
$F''_{m,n}(1) / F_{m,n}(1)$ & -0.125 & -0.103 & 0 & -0.055 & -0.017 & -0.066 & 0.010 & -0.006
\st\\[1ex] \hline
\hline
\end{tabular}
\end{center}
\caption{Table showing the value of $F''_{m,n}(1)/F_{m,n}(1)$ for different $(m,n)$. 
\label{t1}}
\end{table}

Another crude test for determining how good the interpolation formul\ae\ is 
its smoothness. This in turn can be determined
by computing  $F_{m,n}''(g)/F_{m,n}(g)$ around the matching
region $g\sim 1$. In Table \ref{t1} we have shown the value
of $F_{m,n}''(1)/F_{m,n}(1)$ for various interpolating
functions. As we can see, this ratio is largest for the functions $F_{0,0}$ and $F_{0,1}$ --
precisely the two functions whose deviation from $F_{3,1}$ is maximum in Fig.~\ref{f2}.
This gives us indirect indication
that $F_{m,n}$ for $(m,n)$ other than (0,0) and (0,1) are smoother, and hence 
are likely to be
better approximations to the actual result
than $F_{0,0}$ or $F_{0,1}$.

\begin{figure}
\begin{center}
\epsfysize=5cm
\epsfbox{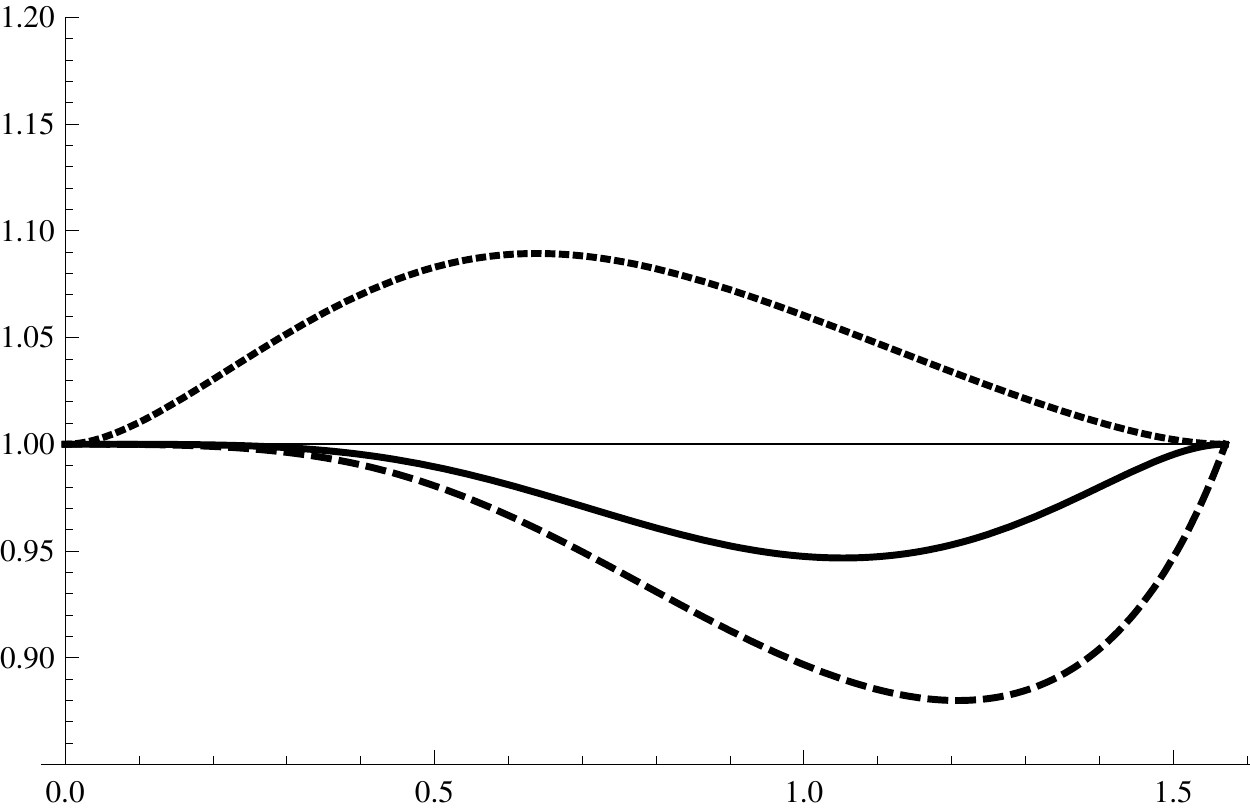}
\end{center}
\caption{Graph of $P_{m,n}(g)/F_{3,1}(g)$ vs. $\tan^{-1}g$ for various $(m,n)$. 
The labels are as follows:   dots for $P_{1,1}$,  dashes for $P_{2,0}$,
continuous thick line for $P_{3,1}$ and
continuous thin line for unity.
\label{fpade}}
\end{figure}

\sectiono{Alternative interpolation formul\ae}  \label{snew}

We can also explore alternative approach to finding the interpolation formula.
A standard  approach is
Pad\'e approximant.\footnote{I wish to thank Barak Kol for suggesting this.}
We look for an interpolation formula of the form
\be \label{epade}
P_{m,n}(g) = g^{1/4} (1 + c_1 g + c_2 g^2 +\cdots c_{p} g^{p} + d_p g^{p+1})^{1/2}
(1 + d_1 g + d_2 g^2 +\cdots d_p g^p)^{-1/2}\, , \quad p = {m+n\over 2}\, ,
\ee
and adjust the $2p$ coefficients $\{c_k\}$ and $\{d_k\}$ to match the weak coupling
expansion to $m$-th order and strong coupling expansion to $n$-th order. Note that
for this approach to work we need $m+n$ to be even. In the special case of $p=0$,
$F_{0,0}$ itself gives  the Pad\'e approximant $P_{0,0}$. 
For our problem, the functions $P_{m,n}$  are
given by
\ben
P_{1,1} &=& g^{1/4} \sqrt{\frac{g^2}{1-2 K_s}+\frac{g}{1-2
   K_s}+1}\Bigg/\sqrt{\frac{g}{1-2 K_s}+1}\, ,
\nonumber \\
P_{2,0} &=& g^{1/4} \sqrt{2 g^2 K_w+2 g K_w+1}\Bigg/ \sqrt{2 g K_w+1}
\nonumber \\
P_{3,1} &=& g^{1/4} \left\{\frac{4 g^3 K_w^2}{4 K_s K_w-2
   K_w+1}+\frac{2 g^2 \left(4 K_s K_w^2+K_w\right)}{4 K_s
   K_w-2 K_w+1}+\frac{2 g K_w}{4 K_s K_w-2
   K_w+1}+1\right\}^{1/2} \nonumber \\
   && \times \left\{\frac{4 g^2 K_w^2}{4 K_s K_w-2
   K_w+1}+\frac{2 g K_w}{4 K_s K_w-2 K_w+1}+1\right\}^{-1/2} \, .
\nonumber \\
\een
Fig.~\ref{fpade} shows the ratios $P_{m,n}/F_{3,1}$ as a function of $g$. As can be
seen from this figure, these ratios also remain within about 12\% unity for all $g$.
In particular $P_{3,1}$ which uses all the available data is within about 5\% of
$F_{3,1}$.

\begin{figure}
\begin{center}
\epsfysize=5cm
\epsfbox{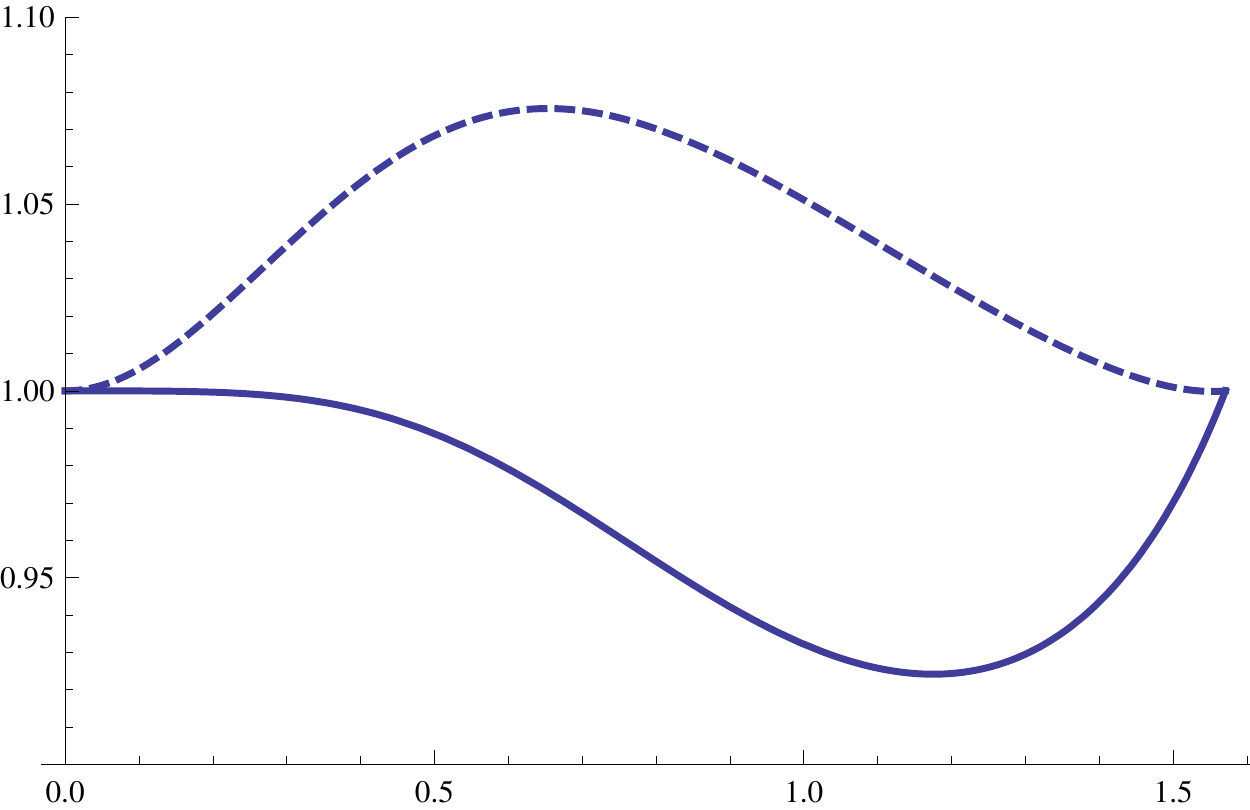}
\end{center}
\caption{Graph of $K_{m,n}(g)/F_{3,1}(g)$ vs. $\tan^{-1}g$ for $(m,n)$
= (1,0) (dashed) and (3,0) (continuous). 
\label{fklein}}
\end{figure}

Another approach is due to Kleinert\cite{kleinert}.\footnote{I wish to
thank Christopher Beem, Leonardo Rastelli and Balt van Rees for drawing my
attention to this method.} This method is designed to generate perturbation expansions
with possibly different powers of coupling at the two ends ({\it e.g.} $g^2$ at
weak coupling end and $g^{-1}$ at the strong coupling end in our case) and
generates interpolating functions which match the weak and the strong coupling
expansions to certain orders. Since this method is somewhat involved we shall not
describe the method here but just quote the results of some approximations.
We denote by $K_{m,n}(g)$ the function that uses weak coupling expansion to
order $g^m$ and strong coupling expansion to order $g^{-n}$ -- in our problem
$m$ will always be odd since the method by construction uses the 
vanishing of the coefficients of odd powers of $g$ at the weak coupling end.
The results are:
\ben
K_{10} &=&\frac{27 g^2+2 \sqrt{81 g^2+4}+4}{\left(\sqrt{81 g^2+4}+2\right)^{3/2}}\, ,
\nonumber \\
K_{30} &=& \left[\left(1.73205 \sqrt{4.7309 g^2+3}+3.\right)^3 \sqrt{\sqrt{14.1927 g^2+9}+3}
\right]^{-1} \nonumber \\ &&
\times \Bigg[
2422.91 g^4+\left(-436.614 \sqrt{4.7309 g^2+3}-1764.56\right) g^2
\nonumber \\ && +\left(-2319.13
   g^2+562.065 \sqrt{4.7309 g^2+3}+2190.43\right) g^2
   \nonumber \\ && +152.735 \sqrt{4.7309
   g^2+3}+264.545
   \Bigg]\, .
\een 
Fig.~\ref{fklein} shows the ratios $K_{10}/F_{3,1}$ and $K_{3,0}/F_{3,1}$ as function
of $g$. Again these remain with 10\% of unity over the entire range of $g$.
Due to the complexity of the algorithm we were not able to determine the
functions $K_{1,1}$ and $K_{3,1}$.

{\bf Acknowledgement:}  I would like to thank Christopher Beem, Rajesh Gopakumar, 
Dileep Jatkar, Barak Kol, Leonardo Rastelli  and Balt van Rees
for discussions and their comments on the original
manuscript.
This work was
supported in part by the J. C. Bose fellowship of 
the Department of Science and Technology, India and the 
project 11-R\&D-HRI-5.02-0304.

\appendix

\sectiono{Testing the algorithm with test functions} \label{s5}

In this appendix  
we shall test the efficiency of the algorithm outlined in \S\ref{s2}
by applying it on some test functions.

\noindent $\bullet$ The first function we consider is
\be 
F(g) = (1+g+g^2)^{1/2}\, .
\ee
The successive approximations $F_{m,n}(g)$ to the function $F(g)$ are taken to be of the form
\be
(1 + a_1 g + \cdots a_m g^m + b_n g^{m+1} +\cdots b_1 g^{m+n} + g^{m+n+1})^{1/(m+n+1)}\, ,
\ee
where the coefficients $a_1,\cdots a_m$ are fixed by requiring that the Taylor series expansion of
$F_{m,n}(g)$ around $g=0$ matches those of $F(g)$ up to order $g^m$, and the coefficients
$b_1, \cdots b_n$ are fixed by requiring that the Taylor series expansion of
$F_{m,n}(g)$ around $g=\infty$ matches those of $F(g)$ up to order $g^{-n+1}$. 
For $m+n+1$ even, $F_{m,n}$ and $F$ are exactly equal since we get
$1 + a_1 g + \cdots a_m g^m + b_n g^{m+1} +\cdots b_1 g^{m+n} + g^{m+n+1}
= (1+g+g^2)^{(m+n+1)/2}$. For odd $m+n+1$ we
get
\ben
&& F_{0,0}(g) = 1+g, \quad F_{1,1}(g) = (1+ 3g/2 + 3g^2/2 + g^3)^{1/3}, \nonumber \\ &&
F_{2,2}(g) = (1+5g/2 + 35 g^2 /8 + 35  g^3/8 + 5 g^4/2 + g^5)^{1/5}, \nn
&& F_{3,3}(g) = (1 + 7 g/2 + 63 g^2/8 + 175 g^3/16 + 175 g^4/16 + 
 63 g^5/8 + 7 g^6/2 + g^7)^{1/7}, \nn
&& F_{4,4}(g) = (1 + 9 g/2 + 99 g^2/8 + 357 g^3/16 + 3843 g^4/128 + 
 3843 g^5/128 + 357 g^6/16 \nn && \qquad \qquad + 99 g^7/8 + 9 g^8/2 + g^9)^{1/9}, 
\een
etc.
Using these we
find that over the entire range of $g$
\ben
&& \left| {F_{0,0}\over F}-1\right| < .155,
\quad \left| {F_{1,1}\over F}-1\right| < .013, \quad
\left| {F_{2,2}\over F}-1\right| < .0021, \nn &&
\left| {F_{3,3}\over F}-1\right| < .00043, \quad \left| {F_{4,4}\over F}-1\right| < .0001, 
\een
etc. Thus the error decreseas for large $m,n$.

\noindent $\bullet$ The second test function we consider is
\be
F(g) = 2 g^{1/4} (1-e^{-1/g})^{1/2} (1+e^{-g})^{-1} + g^{3/4} e^{-1/g}\, .
\ee
Note that there are non-perturbative corrections at both ends and hence we
do not expect to approach the exact result even by going to arbitrary high order.
The approximation $F_{m,n}(g)$ is again taken to be of the form
\be
g^{1/4} (1 + a_1 g + \cdots a_m g^m + b_n g^{m+1} +\cdots b_1 g^{m+n} + g^{m+n+1})^{1/2(m+n+1)}\, .
\ee
The successive approximations are given by
\ben
F_{0,0}(g) &=& g^{1/4} (1+g)^{1/2} \nn
F_{0,1}(g) &=& g^{1/4} (1+4g+g^2)^{1/4} \nn
F_{1,0}(g) &=& g^{1/4} (1+2g+g^2)^{1/4} \nn
F_{1,1}(g) &=& g^{1/4} (1 + 3 g + 6 g^2 + g^3)^{1/6} \nn
F_{2,2}(g) &=& g^{1/4} (1 + 5 g + 45 g^2/4 + 45 g^3 + 10 g^4 + g^5)^{1/10} \nn
F_{3,3}(g) &=& g^{1/4} (1 + 7 g + 91 g^2/4 + 539 g^3/12 + 2905 g^4/8 + 91 g^5 + 
 14 g^6 + g^7)^{1/14} \nn
 F_{4,4}(g) &=& g^{1/4} (1 + 9 g + 153 g^2/4 + 405 g^3/4 + 1479 g^4/8 \nn && + 
 97323 g^5/32 + 6519 g^6/8 + 153 g^7 + 18 g^8 + g^9)^{1/18}\, ,
\een
etc. From this we find that over the entire range of $g$
\ben
&& \left| {F_{0,0}\over F}-1\right| < .119, 
\quad \left| {F_{1,1}\over F}-1\right| < .064, \quad \left| {F_{2,1}\over F}-1\right| < .077, \quad 
\left| {F_{2,2}\over F}-1\right| < .037, \nn &&
\left| {F_{3,3}\over F}-1\right| < .043, \quad 
\left| {F_{4,4}\over F}-1\right| < .057
\een
etc. This shows 
that while the $F_{m,n}$'s can take us quite close to the actual result, 
we do not approach arbitrarily close to the exact result by going to higher orders.
The best result is obtained for $F_{2,2}(g)$ which comes 
within 4\% of $F(g)$ over the entire range of $g$.

\noindent $\bullet$  So far the functions we have analyzed have all the $a_i$'s and $b_i$'s
positive. We shall now give the example of a function that has some of these 
coefficients negative but small enough so that the approximation scheme used here still
takes us sufficiently close to the exact function. We consider 
\be 
F(g) = (1- g/5 +g^2)^{1/2}\, .
\ee
The $F_{m,n}$'s are taken to be
\be
(1 + a_1 g + \cdots a_m g^m + b_n g^{m+1} +\cdots b_1 g^{m+n} + g^{m+n+1})^{1/(m+n+1)}\, .
\ee
In this case again 
for $m+n+1$ even, $F_{m,n}$ and $F$ are exactly equal. For odd $m+n+1$ we
get
\ben
&& F_{0,0}(g) = 1+g, \quad F_{1,1}(g) = (1 - 0.3 g - 0.3 g^2 + g^3)^{1/3}, \nonumber \\ &&
F_{2,2}(g) = (1 - 0.5 g + 2.575 g^2 + 2.575 g^3 - 0.5 g^4 + g^5)^{1/5}, \nn
&& F_{3,3}(g) = (1 - 0.7 g + 3.675 g^2 - 1.7675 g^3 - 1.7675 g^4 + 3.675 g^5 - 
 0.7 g^6 + g^7)^{1/7}, \nn
&& F_{4,4}(g) = (1 - 0.9 g + 4.815 g^2 - 3.2025 g^3 + 8.66644 g^4 + 8.66644 g^5 - 
 3.2025 g^6  \nn && \qquad \qquad + 4.815 g^7 - 0.9 g^8 + g^9)^{1/9}, 
\een
etc.
Using these we
find that over the entire range of $g$
\ben
&& \left| {F_{0,0}\over F}-1\right| < .5,
\quad \left| {F_{1,1}\over F}-1\right| < .17, \quad
\left| {F_{2,2}\over F}-1\right| < .08, \nn &&
\left| {F_{3,3}\over F}-1\right| < .08, \quad \left| {F_{4,4}\over F}-1\right| < .05, 
\een
etc. However the error does not reduce uniformly; instead it
fluctuates around the 5\% mark as we go to higher order. Since the sign of the
error fluctuates, we can considerably reduce the size of the error by averaging the result
for different values of $m,n$.

\noindent $\bullet$ Finally we consider the function
\be
F(g) = \int_{-\infty}^\infty dx\, e^{-x^2/2 - g^2 x^4}\, .
\ee
The expansion around $g=0$ is known to be asymptotic. Since the function goes as
$\sqrt{2\pi}$ as $g\to 0$ and as $(4g)^{-1/2} \Gamma(1/4)$ as $g\to\infty$ we take
\ben
F_{m,n}(g)& = &\sqrt{2\pi} \,
\bigg\{1 + a_1 g + \cdots a_m g^m + b_n g^{m+1} +\cdots b_1 g^{m+n} \nn
&& \qquad + (8\pi)^{(m+n+1)} 
\Gamma(1/4)^{-2(m+n+1)} \,
g^{m+n+1}\bigg\}^{-1/\{2(m+n+1)\}}\, ,
\een
and adjust the coefficients $a_k$ and $b_k$ so as to reproduce the series expansion
around $g=0$ and $g=\infty$. Note that in this case the weak coupling expansion is in
powers of $g^2$ whereas the strong coupling expansion is in powers of $1/g$, exactly as
in our case.
We find that $F_{m,n} (g)$ approaches quite close to $F(g)$ for large enough $(m,n)$.
For example 
\be 
F_{4,4}(g) = \sqrt{2\, \pi} \bigg\{
1 + 54\, g^2 + 594\, g^4 + 780.788 \, g^5 + 1294.34 \, g^6 + 1475.35 \, g^7 + 
 1038.59 \, g^8 + 341.428 \, g^9\bigg\}^{- 1/18}
\ee
 agrees with $F(g)$ within 0.4\%
over the entire range of $g$.

\sectiono{Determination of the normalization constant $\NN$} \label{sa}

\begin{figure}
\begin{center}

\vskip -.5in

\def\JPicScale{0.8}
\ifx\JPicScale\undefined\def\JPicScale{1}\fi
\unitlength \JPicScale mm
\begin{picture}(120,90)(0,0)
\linethickness{1mm}
\put(20,60){\line(1,0){100}}
\put(120,60){\vector(1,0){0.12}}
\linethickness{0.3mm}
\multiput(40,60)(0,-0.5){1}{\line(0,-1){0.5}}
\multiput(40,59.5)(0.01,-0.5){1}{\line(0,-1){0.5}}
\multiput(40.02,59)(0.02,-0.5){1}{\line(0,-1){0.5}}
\multiput(40.04,58.5)(0.03,-0.5){1}{\line(0,-1){0.5}}
\multiput(40.07,58)(0.04,-0.5){1}{\line(0,-1){0.5}}
\multiput(40.1,57.5)(0.05,-0.5){1}{\line(0,-1){0.5}}
\multiput(40.15,57)(0.05,-0.5){1}{\line(0,-1){0.5}}
\multiput(40.21,56.5)(0.06,-0.5){1}{\line(0,-1){0.5}}
\multiput(40.27,56)(0.07,-0.5){1}{\line(0,-1){0.5}}
\multiput(40.34,55.51)(0.08,-0.5){1}{\line(0,-1){0.5}}
\multiput(40.42,55.01)(0.09,-0.49){1}{\line(0,-1){0.49}}
\multiput(40.51,54.52)(0.1,-0.49){1}{\line(0,-1){0.49}}
\multiput(40.6,54.02)(0.1,-0.49){1}{\line(0,-1){0.49}}
\multiput(40.71,53.53)(0.11,-0.49){1}{\line(0,-1){0.49}}
\multiput(40.82,53.05)(0.12,-0.49){1}{\line(0,-1){0.49}}
\multiput(40.94,52.56)(0.13,-0.48){1}{\line(0,-1){0.48}}
\multiput(41.07,52.07)(0.14,-0.48){1}{\line(0,-1){0.48}}
\multiput(41.2,51.59)(0.14,-0.48){1}{\line(0,-1){0.48}}
\multiput(41.35,51.11)(0.15,-0.48){1}{\line(0,-1){0.48}}
\multiput(41.5,50.63)(0.16,-0.47){1}{\line(0,-1){0.47}}
\multiput(41.66,50.16)(0.17,-0.47){1}{\line(0,-1){0.47}}
\multiput(41.83,49.69)(0.18,-0.47){1}{\line(0,-1){0.47}}
\multiput(42,49.22)(0.09,-0.23){2}{\line(0,-1){0.23}}
\multiput(42.19,48.75)(0.1,-0.23){2}{\line(0,-1){0.23}}
\multiput(42.38,48.29)(0.1,-0.23){2}{\line(0,-1){0.23}}
\multiput(42.58,47.83)(0.1,-0.23){2}{\line(0,-1){0.23}}
\multiput(42.79,47.37)(0.11,-0.23){2}{\line(0,-1){0.23}}
\multiput(43,46.92)(0.11,-0.22){2}{\line(0,-1){0.22}}
\multiput(43.22,46.47)(0.11,-0.22){2}{\line(0,-1){0.22}}
\multiput(43.45,46.02)(0.12,-0.22){2}{\line(0,-1){0.22}}
\multiput(43.69,45.58)(0.12,-0.22){2}{\line(0,-1){0.22}}
\multiput(43.94,45.14)(0.13,-0.22){2}{\line(0,-1){0.22}}
\multiput(44.19,44.71)(0.13,-0.21){2}{\line(0,-1){0.21}}
\multiput(44.45,44.28)(0.13,-0.21){2}{\line(0,-1){0.21}}
\multiput(44.71,43.86)(0.14,-0.21){2}{\line(0,-1){0.21}}
\multiput(44.99,43.44)(0.14,-0.21){2}{\line(0,-1){0.21}}
\multiput(45.27,43.02)(0.14,-0.21){2}{\line(0,-1){0.21}}
\multiput(45.55,42.61)(0.15,-0.2){2}{\line(0,-1){0.2}}
\multiput(45.85,42.2)(0.1,-0.13){3}{\line(0,-1){0.13}}
\multiput(46.15,41.8)(0.1,-0.13){3}{\line(0,-1){0.13}}
\multiput(46.46,41.41)(0.1,-0.13){3}{\line(0,-1){0.13}}
\multiput(46.77,41.02)(0.11,-0.13){3}{\line(0,-1){0.13}}
\multiput(47.09,40.63)(0.11,-0.13){3}{\line(0,-1){0.13}}
\multiput(47.42,40.25)(0.11,-0.12){3}{\line(0,-1){0.12}}
\multiput(47.75,39.88)(0.11,-0.12){3}{\line(0,-1){0.12}}
\multiput(48.09,39.51)(0.12,-0.12){3}{\line(0,-1){0.12}}
\multiput(48.44,39.14)(0.12,-0.12){3}{\line(0,-1){0.12}}
\multiput(48.79,38.79)(0.12,-0.12){3}{\line(1,0){0.12}}
\multiput(49.14,38.44)(0.12,-0.12){3}{\line(1,0){0.12}}
\multiput(49.51,38.09)(0.12,-0.11){3}{\line(1,0){0.12}}
\multiput(49.88,37.75)(0.12,-0.11){3}{\line(1,0){0.12}}
\multiput(50.25,37.42)(0.13,-0.11){3}{\line(1,0){0.13}}
\multiput(50.63,37.09)(0.13,-0.11){3}{\line(1,0){0.13}}
\multiput(51.02,36.77)(0.13,-0.1){3}{\line(1,0){0.13}}
\multiput(51.41,36.46)(0.13,-0.1){3}{\line(1,0){0.13}}
\multiput(51.8,36.15)(0.13,-0.1){3}{\line(1,0){0.13}}
\multiput(52.2,35.85)(0.2,-0.15){2}{\line(1,0){0.2}}
\multiput(52.61,35.55)(0.21,-0.14){2}{\line(1,0){0.21}}
\multiput(53.02,35.27)(0.21,-0.14){2}{\line(1,0){0.21}}
\multiput(53.44,34.99)(0.21,-0.14){2}{\line(1,0){0.21}}
\multiput(53.86,34.71)(0.21,-0.13){2}{\line(1,0){0.21}}
\multiput(54.28,34.45)(0.21,-0.13){2}{\line(1,0){0.21}}
\multiput(54.71,34.19)(0.22,-0.13){2}{\line(1,0){0.22}}
\multiput(55.14,33.94)(0.22,-0.12){2}{\line(1,0){0.22}}
\multiput(55.58,33.69)(0.22,-0.12){2}{\line(1,0){0.22}}
\multiput(56.02,33.45)(0.22,-0.11){2}{\line(1,0){0.22}}
\multiput(56.47,33.22)(0.22,-0.11){2}{\line(1,0){0.22}}
\multiput(56.92,33)(0.23,-0.11){2}{\line(1,0){0.23}}
\multiput(57.37,32.79)(0.23,-0.1){2}{\line(1,0){0.23}}
\multiput(57.83,32.58)(0.23,-0.1){2}{\line(1,0){0.23}}
\multiput(58.29,32.38)(0.23,-0.1){2}{\line(1,0){0.23}}
\multiput(58.75,32.19)(0.23,-0.09){2}{\line(1,0){0.23}}
\multiput(59.22,32)(0.47,-0.18){1}{\line(1,0){0.47}}
\multiput(59.69,31.83)(0.47,-0.17){1}{\line(1,0){0.47}}
\multiput(60.16,31.66)(0.47,-0.16){1}{\line(1,0){0.47}}
\multiput(60.63,31.5)(0.48,-0.15){1}{\line(1,0){0.48}}
\multiput(61.11,31.35)(0.48,-0.14){1}{\line(1,0){0.48}}
\multiput(61.59,31.2)(0.48,-0.14){1}{\line(1,0){0.48}}
\multiput(62.07,31.07)(0.48,-0.13){1}{\line(1,0){0.48}}
\multiput(62.56,30.94)(0.49,-0.12){1}{\line(1,0){0.49}}
\multiput(63.05,30.82)(0.49,-0.11){1}{\line(1,0){0.49}}
\multiput(63.53,30.71)(0.49,-0.1){1}{\line(1,0){0.49}}
\multiput(64.02,30.6)(0.49,-0.1){1}{\line(1,0){0.49}}
\multiput(64.52,30.51)(0.49,-0.09){1}{\line(1,0){0.49}}
\multiput(65.01,30.42)(0.5,-0.08){1}{\line(1,0){0.5}}
\multiput(65.51,30.34)(0.5,-0.07){1}{\line(1,0){0.5}}
\multiput(66,30.27)(0.5,-0.06){1}{\line(1,0){0.5}}
\multiput(66.5,30.21)(0.5,-0.05){1}{\line(1,0){0.5}}
\multiput(67,30.15)(0.5,-0.05){1}{\line(1,0){0.5}}
\multiput(67.5,30.1)(0.5,-0.04){1}{\line(1,0){0.5}}
\multiput(68,30.07)(0.5,-0.03){1}{\line(1,0){0.5}}
\multiput(68.5,30.04)(0.5,-0.02){1}{\line(1,0){0.5}}
\multiput(69,30.02)(0.5,-0.01){1}{\line(1,0){0.5}}
\multiput(69.5,30)(0.5,-0){1}{\line(1,0){0.5}}
\multiput(70,30)(0.5,0){1}{\line(1,0){0.5}}
\multiput(70.5,30)(0.5,0.01){1}{\line(1,0){0.5}}
\multiput(71,30.02)(0.5,0.02){1}{\line(1,0){0.5}}
\multiput(71.5,30.04)(0.5,0.03){1}{\line(1,0){0.5}}
\multiput(72,30.07)(0.5,0.04){1}{\line(1,0){0.5}}
\multiput(72.5,30.1)(0.5,0.05){1}{\line(1,0){0.5}}
\multiput(73,30.15)(0.5,0.05){1}{\line(1,0){0.5}}
\multiput(73.5,30.21)(0.5,0.06){1}{\line(1,0){0.5}}
\multiput(74,30.27)(0.5,0.07){1}{\line(1,0){0.5}}
\multiput(74.49,30.34)(0.5,0.08){1}{\line(1,0){0.5}}
\multiput(74.99,30.42)(0.49,0.09){1}{\line(1,0){0.49}}
\multiput(75.48,30.51)(0.49,0.1){1}{\line(1,0){0.49}}
\multiput(75.98,30.6)(0.49,0.1){1}{\line(1,0){0.49}}
\multiput(76.47,30.71)(0.49,0.11){1}{\line(1,0){0.49}}
\multiput(76.95,30.82)(0.49,0.12){1}{\line(1,0){0.49}}
\multiput(77.44,30.94)(0.48,0.13){1}{\line(1,0){0.48}}
\multiput(77.93,31.07)(0.48,0.14){1}{\line(1,0){0.48}}
\multiput(78.41,31.2)(0.48,0.14){1}{\line(1,0){0.48}}
\multiput(78.89,31.35)(0.48,0.15){1}{\line(1,0){0.48}}
\multiput(79.37,31.5)(0.47,0.16){1}{\line(1,0){0.47}}
\multiput(79.84,31.66)(0.47,0.17){1}{\line(1,0){0.47}}
\multiput(80.31,31.83)(0.47,0.18){1}{\line(1,0){0.47}}
\multiput(80.78,32)(0.23,0.09){2}{\line(1,0){0.23}}
\multiput(81.25,32.19)(0.23,0.1){2}{\line(1,0){0.23}}
\multiput(81.71,32.38)(0.23,0.1){2}{\line(1,0){0.23}}
\multiput(82.17,32.58)(0.23,0.1){2}{\line(1,0){0.23}}
\multiput(82.63,32.79)(0.23,0.11){2}{\line(1,0){0.23}}
\multiput(83.08,33)(0.22,0.11){2}{\line(1,0){0.22}}
\multiput(83.53,33.22)(0.22,0.11){2}{\line(1,0){0.22}}
\multiput(83.98,33.45)(0.22,0.12){2}{\line(1,0){0.22}}
\multiput(84.42,33.69)(0.22,0.12){2}{\line(1,0){0.22}}
\multiput(84.86,33.94)(0.22,0.13){2}{\line(1,0){0.22}}
\multiput(85.29,34.19)(0.21,0.13){2}{\line(1,0){0.21}}
\multiput(85.72,34.45)(0.21,0.13){2}{\line(1,0){0.21}}
\multiput(86.14,34.71)(0.21,0.14){2}{\line(1,0){0.21}}
\multiput(86.56,34.99)(0.21,0.14){2}{\line(1,0){0.21}}
\multiput(86.98,35.27)(0.21,0.14){2}{\line(1,0){0.21}}
\multiput(87.39,35.55)(0.2,0.15){2}{\line(1,0){0.2}}
\multiput(87.8,35.85)(0.13,0.1){3}{\line(1,0){0.13}}
\multiput(88.2,36.15)(0.13,0.1){3}{\line(1,0){0.13}}
\multiput(88.59,36.46)(0.13,0.1){3}{\line(1,0){0.13}}
\multiput(88.98,36.77)(0.13,0.11){3}{\line(1,0){0.13}}
\multiput(89.37,37.09)(0.13,0.11){3}{\line(1,0){0.13}}
\multiput(89.75,37.42)(0.12,0.11){3}{\line(1,0){0.12}}
\multiput(90.12,37.75)(0.12,0.11){3}{\line(1,0){0.12}}
\multiput(90.49,38.09)(0.12,0.12){3}{\line(1,0){0.12}}
\multiput(90.86,38.44)(0.12,0.12){3}{\line(1,0){0.12}}
\multiput(91.21,38.79)(0.12,0.12){3}{\line(0,1){0.12}}
\multiput(91.56,39.14)(0.12,0.12){3}{\line(0,1){0.12}}
\multiput(91.91,39.51)(0.11,0.12){3}{\line(0,1){0.12}}
\multiput(92.25,39.88)(0.11,0.12){3}{\line(0,1){0.12}}
\multiput(92.58,40.25)(0.11,0.13){3}{\line(0,1){0.13}}
\multiput(92.91,40.63)(0.11,0.13){3}{\line(0,1){0.13}}
\multiput(93.23,41.02)(0.1,0.13){3}{\line(0,1){0.13}}
\multiput(93.54,41.41)(0.1,0.13){3}{\line(0,1){0.13}}
\multiput(93.85,41.8)(0.1,0.13){3}{\line(0,1){0.13}}
\multiput(94.15,42.2)(0.15,0.2){2}{\line(0,1){0.2}}
\multiput(94.45,42.61)(0.14,0.21){2}{\line(0,1){0.21}}
\multiput(94.73,43.02)(0.14,0.21){2}{\line(0,1){0.21}}
\multiput(95.01,43.44)(0.14,0.21){2}{\line(0,1){0.21}}
\multiput(95.29,43.86)(0.13,0.21){2}{\line(0,1){0.21}}
\multiput(95.55,44.28)(0.13,0.21){2}{\line(0,1){0.21}}
\multiput(95.81,44.71)(0.13,0.22){2}{\line(0,1){0.22}}
\multiput(96.06,45.14)(0.12,0.22){2}{\line(0,1){0.22}}
\multiput(96.31,45.58)(0.12,0.22){2}{\line(0,1){0.22}}
\multiput(96.55,46.02)(0.11,0.22){2}{\line(0,1){0.22}}
\multiput(96.78,46.47)(0.11,0.22){2}{\line(0,1){0.22}}
\multiput(97,46.92)(0.11,0.23){2}{\line(0,1){0.23}}
\multiput(97.21,47.37)(0.1,0.23){2}{\line(0,1){0.23}}
\multiput(97.42,47.83)(0.1,0.23){2}{\line(0,1){0.23}}
\multiput(97.62,48.29)(0.1,0.23){2}{\line(0,1){0.23}}
\multiput(97.81,48.75)(0.09,0.23){2}{\line(0,1){0.23}}
\multiput(98,49.22)(0.18,0.47){1}{\line(0,1){0.47}}
\multiput(98.17,49.69)(0.17,0.47){1}{\line(0,1){0.47}}
\multiput(98.34,50.16)(0.16,0.47){1}{\line(0,1){0.47}}
\multiput(98.5,50.63)(0.15,0.48){1}{\line(0,1){0.48}}
\multiput(98.65,51.11)(0.14,0.48){1}{\line(0,1){0.48}}
\multiput(98.8,51.59)(0.14,0.48){1}{\line(0,1){0.48}}
\multiput(98.93,52.07)(0.13,0.48){1}{\line(0,1){0.48}}
\multiput(99.06,52.56)(0.12,0.49){1}{\line(0,1){0.49}}
\multiput(99.18,53.05)(0.11,0.49){1}{\line(0,1){0.49}}
\multiput(99.29,53.53)(0.1,0.49){1}{\line(0,1){0.49}}
\multiput(99.4,54.02)(0.1,0.49){1}{\line(0,1){0.49}}
\multiput(99.49,54.52)(0.09,0.49){1}{\line(0,1){0.49}}
\multiput(99.58,55.01)(0.08,0.5){1}{\line(0,1){0.5}}
\multiput(99.66,55.51)(0.07,0.5){1}{\line(0,1){0.5}}
\multiput(99.73,56)(0.06,0.5){1}{\line(0,1){0.5}}
\multiput(99.79,56.5)(0.05,0.5){1}{\line(0,1){0.5}}
\multiput(99.85,57)(0.05,0.5){1}{\line(0,1){0.5}}
\multiput(99.9,57.5)(0.04,0.5){1}{\line(0,1){0.5}}
\multiput(99.93,58)(0.03,0.5){1}{\line(0,1){0.5}}
\multiput(99.96,58.5)(0.02,0.5){1}{\line(0,1){0.5}}
\multiput(99.98,59)(0.01,0.5){1}{\line(0,1){0.5}}
\multiput(100,59.5)(0,0.5){1}{\line(0,1){0.5}}

\put(25,65){\makebox(0,0)[cc]{$k$}}

\put(110,65){\makebox(0,0)[cc]{$k$}}

\put(70,25){\makebox(0,0)[cc]{$\ell$}}

\put(70,65){\makebox(0,0)[cc]{$k-\ell$}}
\end{picture}
\end{center}

\vskip -.8in

\caption{A Feynman diagram of low energy effective field theory contributing to mass
renormalization. The thick line represents the propagator of the massive scalar in the
spinor representation of SO(32) and the thin line represents a gauge boson propagator. \label{ff1}}
\end{figure}
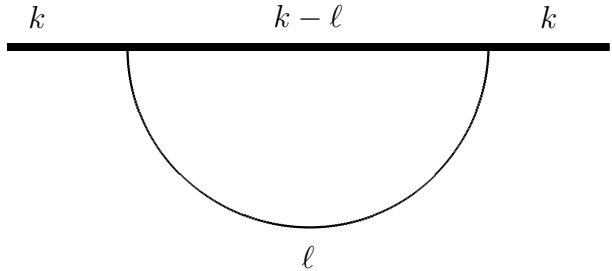

In this appendix we shall determine the normalization constant $\NN$ appearing in 
\refb{egendm} by comparing  this with the corresponding expression for $\delta M^2$ in
the field theory limit. Let $\chi_s$ denote a massive scalar in the spinor representation
of the SO(32) gauge group. Then the action of the low energy effective field theory
describing the SO(32) gauge fields and
their coupling to the scalars $\chi_s$  is given by
\be\int d^{10} x \bigg[ - {1\over 16 (g_H)^2 } Tr_V (F_{\mu\nu} F^{\mu\nu}) 
- D_\mu \chi_s^* D^\mu\chi_s - M^2 \chi_s^* \chi_s
\bigg]\, .
\ee
Note that the gauge field action has been normalized in accordance with \refb{e1het} 
by setting $G^H_{\mu\nu}=\eta_{\mu\nu}$ and
the normalization of the kinetic term for $\chi_s$ is standard.
We shall compute the mass renormalization of $\chi_s$ due to the one loop diagram of
Fig.~\ref{ff1}. More specifically we shall compute the contribution from the region of loop
momentum integration where the momentum $\ell$ carried by the vector fields is
small. In this limit we essentially compute the correction to the mass due to the Coulomb
field of the $\chi$ particle and the result is independent of the spin of $\chi$. Indeed it
is easy to verify that our results remain unchanged if instead of $\chi$ we use another
field with a different spin {\it e.g.} a Dirac field. This is important since the SO(32) spinor
states for which we need to compute the mass renormalization in string theory are not
scalars but transform in the {\bf 84+44} representation of the little group $SO(9)$.

In order to compute the contribution from the graph shown in
Fig~\ref{ff1} we first need to choose a convention for the SO(32) generators
$T^a$. In the vector representation we choose them to be $32\times 32$ matrices whose
$mn$ element is $i$, $nm$ element is $-i$ and whose other elements are zero.
Collection of all such matrices for $1\le m<n\le 32$ form the basis of SO(32) generators.
With this convention we have $Tr_V(T^a T^b) = 2 \delta_{ab}$ and hence the gauge kinetic
term is normalized as $-{1\over 8} F^a_{\mu\nu} F^{a\mu\nu}$. As a result the gauge
field propagator of momentum $\ell$ in the Feynman gauge will be of the form 
$-2 \, i \, \delta_{ab} g_{\mu\nu} / (\ell^2
-i\epsilon)$.  The coupling of the gauge field to $\chi$ in the Lgrangian density 
is of the form $i A^{a\mu} (T^a)_{ss'}
(\chi_s^* \p_\mu \chi_{s'} - \p_\mu \chi_s^* \chi_{s'})$ from which we can read out the vertex. This
gives the contribution to $\delta M^2$ to be
\be \label{egaugecont}
-i \delta M^2 = 2\, (T^a T^a)_{spinor} \, 
\int{d^{10}\ell\over (2\pi)^{10}} {1\over \ell^2 - i\epsilon}
{1\over (k-\ell)^2 + M^2 - i\epsilon} (2 k_\mu - \ell_\mu) (2 k^\mu - \ell^\mu)
\ee
where the momentum integration is still written in the Lorentzian space, and
$(T^aT^a)_{spinor}$ denotes the eigenvalue of $T^aT^a$ on the spinor
representation. With the normalization of $T^a$ we have chosen, 
each $T^a$ has eigenvalue $\pm 1/2$ on the spinor representation and hence,
after summing over the 496 generators of SO(32), 
$T^a T^a$ will have eigenvalue 124 on the spinor representation. 

We shall
analyze the contribution to \refb{egaugecont}
from the region of integration where $\ell$ is small so that
the low energy effective field theory description makes sense. In this case we can replace
the vertex factor $(2 k_\mu - \ell_\mu) (2 k^\mu - \ell^\mu)$ by $4 k_\mu k^\mu = - 4(k_0)^2$ and
take it out of the integral. Making the Wick rotation $\ell^0\to i \ell^0$ and denoting by
$\ell_E$ the Euclidean momentum we get
\be 
\delta M_{gauge}^2 = 8\, (k_0)^2 \times  124 \times
\int{d^{10}\ell_E\over (2\pi)^{10}} {1\over \ell_E^2}
{1\over (\ell_E-k)^2 + M^2}\, .
\ee
In order to make connection with the string theory result \refb{egendm} we now need to
express the propagators in the Schwinger proper time formalism. We write
\be 
\delta M_{gauge}^2 = 992\, (k_0)^2 \, \pi^2 \int{d^{10}\ell_E\over (2\pi)^{10}} \int_0^\infty d y_2 
\int_0^\infty d z_2 \, e^{-\pi y_2 \ell_E^2 - \pi z_2 ((\ell_E-k)^2 + M^2)}
\, .
\ee
Changing variable to $\tau_2=y_2+z_2$ and $z_2$ we get
\ben \label{efingauge}
\delta M_{gauge}^2 &=& 992\, (k_0)^2 \, \pi^2 \int{d^{10}\ell_E\over (2\pi)^{10}} \int_0^\infty d \tau_2 
\int_0^{\tau_2} d z_2 \, e^{-\pi (\tau_2-z_2) \ell_E^2 - \pi z_2 ((\ell_E-k)^2 + M^2)}\nn
&=& 992\, (k_0)^2 \, \pi^2 \int{d^{10}\ell_E\over (2\pi)^{10}} \int_0^\infty d \tau_2 
\int_0^{\tau_2} d z_2 \, e^{-\pi \tau_2 (\ell_E - z_2 k/\tau_2)^2
+ \pi z_2^2 k^2 / \tau_2 - \pi z_2 (k^2 + M^2)}\nn
&=& 992\, (k_0)^2 \, \pi^2 (2\pi)^{-10}  \int_0^\infty d \tau_2 
\int_0^{\tau_2} d z_2 \, (\tau_2)^{-5}\, e^{-4\pi z_2^2 / \tau_2}
\, ,
\een
where in the last step we have carried out the integration over $\ell_E$ and have also
used the on-shell condition $k^2 = - M^2 =-4$ in the exponent.

We shall now try to reproduce the same integral from the string theory result 
\refb{egendm}. We shall focus on the region of integration where both $z_2$ and
$\tau_2$ are large but $z_2$ remains small compared to $\tau_2$. Since the integrand in
\refb{egendm} is invariant under $z\to \tau-z$, there will be an identical contribution from the region
where $(\tau_2-z_2)$ is small compared to $\tau_2$, and the effect of this will be to
double the contribution from the $z_2 << \tau_2$ region.\footnote{Physically $z\to \tau-z$
transformation exchanges the role of the two arms of the torus between the points 0 and
$z$.}  Now since in computing the field theory contribution we examined only 
the graphs with bosonic intermediate states, we must do the same in string theory. This
corresponds to restricting the sum over spin structure $\nu$ to $00$ and
$01$ sectors only.  We now use the following approximations to the various
factors in \refb{egendm} for large $\tau_2$, $z_2$:\footnote{Note that the leading term
in the integrand has a stronger singularity in $\bar \tau$ compared to $\tau$ and hence in the
expansion in $e^{-2\pi i\bar\tau}$ and $e^{-2\pi i \bar z}$ we must keep more terms than in
their holomorphic counterparts. The general rule is that while expanding the holomorphic terms
we only keep the leading terms except in those inside the sum over the spin structure $\nu$
where we have to keep the first subleading terms. On the other hand for the anti-holomorphic terms
we need to keep up to subleading terms of order $e^{-2\pi i \bar\tau}$,}
\ben 
\vt_{00}(z)^4 &\simeq& 1 + 4\, e^{\pi i \tau - 2\pi i z} + 4\, e^{\pi i \tau + 2\pi i z} + \cdots \nn
\vt_{01}(z)^4 &\simeq& 1 - 4\, e^{\pi i \tau - 2\pi i z} - 4\, e^{\pi i \tau + 2\pi i z} + \cdots \nn
\vt_{11}(z)^{-1} &\simeq& -i\, e^{-\pi i \tau/4} e^{\pi i z} (1 + \cdots) \nn
\eta(\tau)^{-1} &\simeq& e^{-\pi i \tau / 12} \, (1+\cdots)  \nn
\overline{\vt_{00}(z/2)}^{16} &\simeq& 1 + 16\,  e^{-\pi i \bar\tau + \pi i \bar z} + 16 \, e^{-\pi i 
\bar\tau - \pi i \bar z}
+ 120 e^{-2\pi i \bar\tau + 2\pi i \bar z} + \cdots \nn
\overline{\vt_{01}(z/2)}^{16} &\simeq& 1 - 16\,  e^{-\pi i \bar\tau + \pi i \bar z} - 16 \, e^{-\pi i 
\bar\tau - \pi i \bar z}  
+ 120 e^{-2\pi i \bar\tau + 2\pi i \bar z} + \cdots \nn 
\overline{\vt_{10}(z/2)}^{16} &\simeq& 0 \nn
\overline{\vt_{11}(z/2)}^{16} &\simeq& 0 \nn
(\overline{\vt_{11}(z)})^{-1} &\simeq& i\, e^{\pi i \bar\tau/4} e^{-\pi i \bar z} (1 - e^{-2\pi i \bar z}
- e^{-2\pi i (\bar\tau - \bar z)} + \cdots)^{-1} \nn
& \simeq & e^{\pi i \bar\tau/4} e^{-\pi i \bar z}  \,
(1 +  \, e^{-2\pi i \bar z} + e^{-4\pi i \bar z} + \cdots 
+ e^{2\pi i (\bar\tau - \bar z)} + \cdots )\, , \nn
(\overline{\eta(\tau)})^{-1} &\simeq& e^{\pi i \bar \tau/12}  (1 + e^{-2\pi i \bar\tau} + \cdots) \, . 
\een
Substituting these into \refb{egendm}, and taking into account the extra factor of 2 due to $z\to 
\tau-z$ symmetry, we get the string theory result for $\delta M^2$ from the region where $z_2$ and
$\tau_2$ are large:
\ben \label{estring}
\delta M^2 &\simeq&    8 \times \NN \, (k_0)^2 \, 
\int_{\tau_2>>1}  d^2\tau \int_{1<< z_2 << \tau_2} d^2 z \,  \,e^{2\pi i (\bar\tau - \bar z)} \, \nn &&
\bigg\{ (1 + 4 e^{-2\pi i \bar\tau}+\cdots) (1 + 4 e^{-2\pi i \bar z} + 4 e^{-2\pi i \bar\tau + 2\pi i \bar z}+\cdots) 
( (1 + 120 e^{-2\pi i \bar\tau + 2\pi i \bar z} + \cdots) \bigg\}\nn &&
\{ 1+\cdots \} \, \{ 
 (1 + 2  \, e^{-2\pi i \bar z} + 2\, e^{-2\pi i (\bar\tau - \bar z)} +\cdots ) \,  (1 + 14 e^{-2\pi i \bar\tau} + \cdots) 
 e^{-4\pi \, z_2^2 / \tau_2} \, (\tau_2)^{-5} \, . \nn
\een
This expression has been organized as follows. After factoring out the leading terms inside
each of the curly brackets in \refb{egendm}, we have written inside the three curly brackets
the subleading terms from the terms inside the three curly brackets in \refb{egendm}.
In particular since the leading holomorphic term has no factors of $e^{-2\pi i \tau}$ or
$e^{-2\pi i z}$ we have dropped all the subleading pieces containing factors of $e^{2\pi i \tau}$
or $e^{2\pi i z}$. 

Now for large $\tau_2$ the integrals over $\tau_1$ and $z_1$ run from $-1/2$ to $1/2$ without
any restriction. Carrying out these integrals projects us into those terms which do not have any
factors of $e^{-2\pi i\bar\tau}$ or $e^{-2\pi i\bar z}$. Thus we need to identify such terms in
\refb{estring}. Furthermore since we are only interested in picking up the contributions due to gauge
boson exchange we need to consider only the subleading terms coming from the expansion of the
terms inside the first curly bracket since this is what arose from the 32 left-moving fermions. 
The $e^{-2\pi i (\bar \tau -\bar z)}$ term in this expansion,
which cancels the overall multiplicative factor of $e^{2\pi i (\bar\tau - \bar z)} $, has coefficient
$(120+4)=124$. Thus the gauge boson exchange contribution to \refb{estring} is given by
\be \label{estrfin}
8 \times 124 \, \NN \, (k_0)^2 \, 
\int d^2\tau \int d^2 z \,   e^{-4\pi \, z_2^2 / \tau_2} \, (\tau_2)^{-5} \, .
\ee
Comparing this with the field theory result \refb{efingauge} we now get
\be \label{enresult}
\NN = {992 \over 8\times 124} \pi^2 (2\pi)^{-10} = 2^{-10} \pi^{-8} \, .
\ee

\end{document}